\documentclass[12pt, preprint]{aastex}
\usepackage{epsfig}
\usepackage{rotate}

\renewcommand\ion[2]{#1\,{\sc{\romannumeral #2}}} 

\newcommand\sk[2]{Sk\,{$#1{^\circ}#2$}}
\newcommand{\fuse}{{\it FUSE}}
\newcommand{\hst}{{\it HST}}
\newcommand{\iue}{{\it IUE}}
\newcommand{\vsini}{$v \sin i$}
\newcommand{\vinf}{${v}_\infty$}
\newcommand{\kms}{km\,s$^{-1}$}
\newcommand{\teff}{$T_{\rm eff}$}
\newcommand{\ebv}{$E$(\bv)}
\newcommand{\msunpyr}{${\rm M}_\sun / {\rm yr}$}

\newcommand{\trad}{$\tau_{rad}$}

\newcommand{\mdot}{$\dot{M}$}
\newcommand{\mdotq}{$\dot{M} q$}
\newcommand{\mdotqi}{$\dot{M} q_i$}
\newcommand{\meanqi}{$\langle q_i \rangle$}

\slugcomment{Version of \today}
\shortauthors{Massa et al.}
\shorttitle{{\fuse}\/ Observations of O Star Winds in the LMC}

\received{2002 July 31}
\begin{document}

\title{Constraints on the Ionization Balance of Hot-Star Winds from
      {\fuse} Observations of O Stars in the Large Magellanic Cloud
       \footnote{Based on observations made with the NASA-CNES-CSA 
        Far Ultraviolet Spectroscopic Explorer. {\fuse} is operated 
        for NASA by the Johns Hopkins University under NASA contract 
        NAS5-32985.} 
        }

\author{D.\ Massa}
\affil{SGT Inc., 
       Code 681.0, 
       NASA's Goddard Space Flight Center, 
       Greenbelt, MD 20771. }
\email{massa@taotaomona.gsfc.nasa.gov}

\author{A. W. Fullerton\altaffilmark{2}}
\affil{Dept. of Physics \& Astronomy, 
       University of Victoria, 
       P.O.\ Box 3055, 
       Victoria, BC, V8W 3P6, Canada.}
\altaffiltext{2}{Dept. of Physics \& Astronomy, 
                 The John Hopkins University, 
                 3400 N.\ Charles St., 
                 Baltimore, MD 21218.}
\email{awf@pha.jhu.edu}

\author{G. \ Sonneborn}
\affil{Laboratory for Astronomy \& Solar Physics, 
       Code 681.0
       NASA's Goddard Space Flight Center, 
       Greenbelt, MD 20771.}
\email{george.sonneborn@gsfc.nasa.gov}

\and

\author{J.\ B.\ Hutchings}
\affil{Herzberg Institute of Astrophysics
       National Research Council of Canada,
       5071 West Saanich Road,
       Victoria, BC, V8X 4M6, Canada.}
\email{John.Hutchings@hia.nrc.ca}

\begin{abstract}

We present Far Ultraviolet Spectroscopic Explorer (\fuse) spectra for 25 O 
stars in the Large Magellanic Cloud (LMC).  We analyze wind profiles for 
the resonance lines from {\ion{C}{3}}, {\ion{N}{3}}, {\ion{S}{4}}, 
{\ion{P}{5}}, {\ion{S}{6}}, and {\ion{O}{6}} in the {\fuse} range using 
a Sobolev with Exact Integration (SEI) method.  In addition, the available 
data from either {\iue}\/ or {\hst}\/ for the resonance lines of 
{\ion{Si}{4}}, {\ion{C}{4}}, and {\ion{N}{5}} are also modeled.  Because 
several of the {\fuse} wind lines are unsaturated, the analysis provides 
meaningful optical depths (or, equivalently, mass loss rate times 
ionization fractions, {\mdotq}), as a function of normalized velocity, 
$w=v/${\vinf}.  Ratios of {\mdotq} (which are independent of {\mdot}) 
determine the behavior of the relative ionization as a function of $w$. 
The results demonstrate that, with the exception of {\ion{O}{6}} in all 
stars and {\ion{S}{6}} in the later stars, the ionization in the winds 
shifts toward lower ionization stages at higher $w$ (contrary to the 
expectations of the nebular approximation). This result implies that the 
dominant production mechanism for {\ion{O}{6}} and {\ion{S}{6}} in the 
late O stars differs from the other ions.  

Using the \citet{Vink01} relationship between stellar parameters and 
mass-loss rate, we convert the measurements into mean ionization fractions 
for each ion, $\langle q_i \rangle$.  Because the derived ion fractions 
never exceed unity, we conclude that the derived values of {\mdot} are 
not too small.  However,  $q$({\ion{P}{5}}), which is expected to be the 
dominant stage of ionization in some of these winds, is never greater 
than 0.20.  This implies that either the calculated values of {\mdot} are 
too large, the assumed abundance of phosphorus is too large or the winds 
are strongly clumped.  The implications of each possibility are discussed. 
Correlations between the mean ion fractions and physical parameters such 
as {\teff}, {\vinf} and the mean wind density, $\langle \rho \rangle$ are 
examined.   Two clear relationships emerge.  First, as expected, the mean 
ionization fraction of the lower ions {\ion{C}{3}}, {\ion{N}{3}}, 
{\ion{Si}{4}}, {\ion{S}{4}}) decreases with increasing {\teff}.  Second, 
the mean ion fraction of {\ion{S}{6}} in the latest stars and {\ion{O}{6}} 
in all stars increases with increasing {\vinf}.  This re-affirms the 
notion, first introduced by \citet{Cassinelli79}, that {\ion{O}{6}} is 
produced non-radiatively.

Finally, we discuss specific characteristics of three stars, BI~272, 
BI~208, and {\sk{-67}{166}}.  For BI~272, the ionic species present in its 
wind suggest a much hotter than its available (uncertain) spectral type of 
O7: II-III:.  In the case of BI~208, our inability to fit its observed 
profiles suggests that its wind is not spherically symmetric.  For 
{\sk{-67}{166}}, quantitative measurements of its line strengths confirm 
the suggestion by \citet{Walborn95a} that it is a nitrogen rich O star.  

\end{abstract}


\section{Introduction}

Although the stellar winds of early-type stars are understood to be driven 
by momentum transfer from the underlying stellar radiation field, there
is overwhelming evidence that the ionization conditions in the wind are 
not solely determined by radiative processes.  Initial indications for the 
influence of additional processes came from observations of wind profiles 
for ``super-ions'' like {\ion{N}{5}} and {\ion{O}{6}} by the 
{\it Copernicus}\/ observatory {\citep[see, e.g.][]{Snow76,Walborn96}}.  
Figure~\ref{ions} is a schematic representation of the ionization ranges 
of the ions whose wind lines are analyzed in this paper as well as those 
ions which do not have observable lines.  The dashed vertical line denotes 
the ionization potential of {\ion{He}{2}}.  Ions which lie completely to 
the right of this line are super-ions.  The abundances of super-ions are 
expected to be quite low in all but the hottest stars, since direct 
photoionization of the ground state of the next lower ion should be rare.  
This is because the flux of photons 
sufficiently energetic to create them is strongly suppressed by bound-free 
absorption of He$^+$ in the photosphere.  However, the strength of the 
super-ion wind lines and, especially, the presence of {\ion{O}{6}}, lead 
\citet{Cassinelli79} to predict that hot-star winds must emit X-rays, 
which could then modify the abundances of super-ions via Auger ionization.  
This prediction was subsequently confirmed by observations from the 
{\it Einstein\/} {\citep{Harnden79,Seward79,Chlebowski89}}, {\it ROSAT}\/ 
{\citep{Berghoefer96}} and {\it ASCA}\/ observatories {\citep{Corcoran94}}.  
More recent observations with the spectrometers onboard {\it Chandra}\/ 
and {\it XMM-Newton}\/ indicate the presence of even more highly ionized 
species {\citep[see, e.g.,][]{Schulz00,Kahn01,Waldron01,Cassinelli01}}.

The origin of X-rays in hot-star winds is still an open issue.
Shocks due to the intrinsic line-driven instability {\citep{Owocki88}}
or the formation of large-scale, co-rotating structures {\citep{Cranmer96}} 
or heating of plasma in a magnetic loops {\citep{udDoula02}} are likely 
possibilities, while models invoking the presence of a deep-seated, hot 
corona {\citep[e.g.,][]{Cassinelli79,Waldron84}} are not considered as 
plausible.  Since their origin is so uncertain, X-rays are included in the
current generation of model atmospheres only in an {\it ad hoc}\/ manner 
{\citep[see, e.g.,][]{Hillier98,Pauldrach94,Pauldrach01}}.  Predicted 
ionization equilibria from model atmosphere calculations are further 
complicated by uncertainties in the treatment of line blocking and 
blanketing in the far- and extreme-ultraviolet regions of the spectrum, 
which determine the flux of hard photospheric photons that are available 
to illuminate the wind.  Thus, despite the impressive progress in 
calculating sophisticated {\it ab initio\/} models of moving atmospheres 
reviewed recently by \citet{Kudritzki00}, self-consistent theoretical 
determinations of the prevailing ionization conditions remain elusive.

Theoretical progress is further complicated by the limited guidance 
provided by observations.  Essentially all empirical information 
concerning the ionization of the wind material comes from observations of 
the ultraviolet (UV) resonance lines of ionized metals, which are the most 
sensitive indicators of mass loss.  By modeling the wind profiles of these 
species, it is possible to determine the radial optical depth (or 
equivalently, column density) of the ion as a function of velocity in the 
stellar wind.  Since the degree of excitation for metal ions is extremely 
low, this can be expressed directly in terms of the product of the 
mass-loss rate and ionization fraction, {\mdotq}, for the species 
\citep[e.g.,~][]{Hamann80,Hamann81b,Lamers87,Groenewegen89b, 
Howarth89,Groenewegen91}.  However, the resonance lines of abundant 
elements, and especially the dominant ionization stages of these elements, 
are often saturated, and provide only lower limits on the quantity of the 
ion in the wind. The spectral diagnostics accessible to {\iue}\/ and 
{\hst}\/ -- usually {\ion{N}{5}}~$\lambda\lambda$1239, 1243, 
{\ion{Si}{4}}~$\lambda\lambda$1394, 1403, and 
{\ion{C}{4}}~$\lambda\lambda$1548, 1551 -- are especially problematical, 
since C and N are cosmically abundant and hence frequently saturated.
Furthermore, even when two or more of these lines are unsaturated
{\it ratios\/} of their respective values of {\mdotqi} still depend on 
the abundance of their parent element.  The abundance of C and N can 
differ substantially from nominal values in the atmospheres of OB stars, 
particularly supergiants, due to mixing of material processed through the 
CNO cycle from the interior.  Thus, even the ratios of ion fractions are 
difficult to interpret from these diagnostics, unless the actual 
abundances of C and N have been determined by some means.

In contrast, the far-ultraviolet (FUV; $\sim$905--1215~{\AA}) region
provides a much richer suite of resonance line diagnostics.
As detailed in \S 4, these include
 (a) lines from cosmically abundant (e.g., O, C, N) and
     comparatively rare (e.g., S, P) elements, which are
     less likely to be saturated; 
 (b) lines from species that are expected to dominant (e.g., {\ion{P}{5}}) 
     and trace (e.g, {\ion{C}{3}}) ions; 
 (c) lines from multiple ionization stages of the same element
     (e.g., {\ion{S}{4}}/{\ion{S}{6}}); and
 (d) lines from super-ions (e.g., {\ion{S}{6}} and {\ion{O}{6}}) and 
     low ions (e.g., {\ion{N}{3}}).
Many of the earliest measurements of resonance lines in the winds of
early-type stars were based on FUV observations obtained by 
{\it Copernicus}; see, e.g., \citet{Gathier81} and \citet{Olson81a}.
However, due to the severity of interstellar extinction and limited 
instrumental sensitivity, spectra below 1000~{\AA} were obtained for an 
extremely limited sample of  Galactic O and early B-type stars, most 
notably $\zeta$~Puppis {\citep[O4~In(f);][]{Morton77}} and $\tau$~Scorpii 
{\citep[B0~V;][]{Rogerson77}}.  As a result, these two objects have 
dominated theoretical modeling efforts for the last two decades, 
particularly those aimed at tracing the distribution of the X-rays believed 
responsible for superionization {\citep[e.g.,][]{MacFarlane93}}.

With the launch of the {\it Far Ultraviolet Spectroscopic Explorer} 
({\fuse}\/) in 1999 June, routine access to the FUV region of the spectrum 
has become available once more.  Unlike {\it Copernicus}, {\fuse}\/ is 
sensitive enough to provide high signal-to-noise (S/N), high resolution 
spectra of O-type stars in the Large and Small Magellanic Clouds (LMC and 
SMC).  Furthermore, in contrast to the sensitive, high-resolution FUV 
spectrographs flown in the on shuttle-based missions in the mid-1990s 
(particularly the {\it Berkeley Extreme and Far Ultraviolet Spectrometer}), 
the duration of the {\fuse} mission permits extensive surveys of the hot 
stars in these galaxies.  Access to OB-type stars in the Magellanic Clouds 
provides several distinct advantages over previous work.

First, early-type stars in the LMC and SMC are only lightly reddened, 
especially when compared with Galactic O stars that are faint enough to 
be observed by {\fuse} (i.e., $F_\lambda \le 1 \times 10^{-10}$ 
erg~s$^{-1}$~cm$^{-2}$~{\AA}$^{-1}$ throughout the FUV region).
As a result, their continuum flux distributions are nearly flat over the 
entire FUV waveband, and problems caused by blending of stellar features
with interstellar absorption lines are more manageable than in most
of their Galactic counterparts.  Thus, {\fuse} observations of early-type 
stars in the Magellanic Clouds provide the first systematic assessment 
of the behavior of the important stellar wind diagnostics below 1000~{\AA}, 
particularly the resonance lines of {\ion{S}{6}}, {\ion{P}{4}}, 
{\ion{C}{3}}, and {\ion{N}{3}}.  These trends are described and illustrated 
in the detailed atlas of {\fuse\/} spectra prepared by \citet{Walborn02a}.

Second, since the abundance of metals in the Magellanic Clouds is smaller 
than in the Milky Way, the effect of metallicity on the bulk properties 
of stellar winds can be probed.  The hot stars in each of the Magellanic 
Clouds lie at a common, known distance and are thought to be a members of 
a homogeneous population.  Consequently, the Magellanic Clouds are ideal 
laboratories for studying the physics of line-driven stellar winds.
Initial analyses of {\fuse} spectra have been reported by 
\citet{Bianchi00}, who performed wind-profile fitting for a matched
pair of O supergiants (one from each Cloud) and \citet{Fullerton00}, who 
used model atmospheres to estimate the stellar parameters for the same 
objects.

In this paper, we present the first comprehensive investigation of
the properties of the stellar winds of O-stars in the LMC at FUV 
wavelengths.  In particular, we use measured optical depths in various
resonance lines and assumed mass-loss rates to determine empirical 
ionization fractions, which provide qualitatively new information 
about the ionization balance in the winds of these stars.  \S2 provides 
a description of the program stars, the observations, and the data 
processing, while \S3 describes the wind-profile fitting technique used to 
characterize the properties of the outflows.  Characteristics of the 
resonance lines analyzed are given in \S 4, which also includes a 
discussion of specific problems associated with fitting them.  The results 
of the measurements and their errors are described in \S5 and discussed in 
\S6.  The conclusions are summarized in \S7.

\section{Observations and Data Processing}

\subsection{Program Stars}

The sample of O stars presented here was largely drawn from programs
designed by the {\fuse}\/ Principal Investigator team to study
the stellar content and interstellar medium (ISM) of the LMC.
The sample represents a subset of the data available in late 2001,
which was selected to provide full coverage of the O-type temperature
classes with minimal reddening.  Stars known to be members of complicated 
binary systems were avoided, and stars with bright neighbors that fell 
within the {\fuse}\/ large aperture were also eliminated.  A secondary 
criterion was that the targets should also have spectra covering the UV 
resonance lines in the $1200 \le \lambda \le 1700$~{\AA} region from 
either high dispersion {\iue}, {\hst}\,/\,FOS, or {\hst}\,/\,STIS 
observations.  Archival spectra from one or more of these sources were 
available for all but two targets ({\sk{-67}{69}} and {\sk{-70}{115}}).

Table~{\ref{startab}} lists the fundamental properties of the 25 stars
in our sample.  Successive columns record the designations of the objects 
in  the catalogs of either \citet[][Sk]{Sanduleak70} or 
\citet[][BI]{Brunet75}; commonly used aliases; the spectral classification 
and its source; the terminal velocity ({\vinf}) of the wind as determined 
by the current investigation; the adopted effective temperature ({\teff}) 
from the spectral-type -- {\teff} calibration of \citet{Vacca96};
the computed stellar luminosity; and the adopted mass-loss rate (\mdot).
Table~{\ref{datatab}} lists the optical data ($V$ and \bv) and their
origin, the derived {\ebv}, the observed FUV flux value at the fiducial
wavelength 1150~\AA\, and the identifications of  the {\fuse}\/ 
observations, {\iue}, or {\hst} data sets used. 

The stellar luminosities in Table~{\ref{startab}} were calculated using the 
the \citet{Vacca96} bolometric corrections, $BC$, for the adopted {\teff}, 
the observed $V$ magnitude given in Table~{\ref{datatab}}, corrected for 
extinction (using the spectral type -- {(\bv)$_0$} calibration of 
\citet{FitzGerald} and a total-to-selective extinction ratio of 
$R(V) = A(V)/E(B-V) = 3.1$) and an adopted LMC distance modulus ($DM$) of 
18.52 {\citep{Fitz2002}}.  Deviations from the assumed value of $R(V)$ 
have little effect on the luminosity, since the reddenings are quite 
modest.  

The mass-loss rates in Table~{\ref{startab}} were estimated from the 
adopted stellar parameters by using the relationship derived by 
\citet[][~their eq.~{[15]}]{Vink00}
\begin{equation}
\log \dot{M} = -15.88 +1.576 \log L/L_\sun + \log T_{\rm eff} -\log v_\infty
\end{equation}
These estimates were further adjusted to account for the lower metallicity 
of the LMC by means of the scaling relation
\begin{equation}
   \log \dot{M} = \log \dot{M} + 0.69 \log Z/Z_\sun
\end{equation}
from  \citet{Vink01}, where we adopt $Z/Z_\sun = 0.5$ for the LMC 
{\citep{Rolleston,Welty}}.  This value lies between the extremes 
examined by Vink et al.\ and provides good agreement between their 
predicted mass loss rates and those observed by \citet{Puls96}  (see 
Figure 6 in Vink et al.).  

Insofar as \citet{Vacca96}, \citet{Vink01} and \citet{Puls96} used similar 
temperature calibrations, we expect the estimates of {\mdot} to be 
internally consistent.  Accordingly, the {\em random} uncertainty in 
{\mdot} is dominated by the uncertainties in {\teff}, $\log L/L_\sun$ 
and $\log Z/Z_\sun$.  The effective temperature of a program star cannot 
be known better than half a spectral class, which corresponds to $\sim 
1.5$~kK for a typical O star temperature of 40~kK (Vacca et al.\ 1996).  
This error enters the $\dot{M}$ determination both explicitly and 
implicitly through its effect on the $BC$ which is used to determine the 
luminosity; specifically, $\log L/L_\odot = -0.4(M_V +BC -M_{bol,\odot})$.  
Since $M_V$ is constrained by the adopted distance to the LMC, the only 
error in $\log L/L_\odot$ is due to the temperature uncertainty in their 
$BC$.  A typical half spectral class change in the $BC$ is $-0.1$, which 
translates into $+0.04$ in $\log L/L_\odot$.  Thus, for typical O star 
values, a $\log \dot{M}$ error due to the expected uncertainty in {\teff} 
is $1.576(0.04) - \log (1 + 1500/40000) \sim 0.05$, or 12\%.  Another error 
results from the uncertainty in the LMC distance, which is at least 
$\pm 0.1$ in the distance modulus, which corresponds to an error of $1.576 
\times 0.4 \times 0.1$ in $\log \dot{M}$, or about 15\% in $\dot{M}$.  
Finally, $\log Z/Z_\sun$ is not known to better than 20\%, and this error 
propagates to 14\% in $\dot{M}$.  Now, if we assume that each of these 
errors (due to temperature, LMC distance and abundances) are independent, 
then (as long as the Vink et al.\ relations are exact), the overall 
uncertainty in the derived $\dot{M}$ values should be less than 25\%.

\subsection{Far Ultraviolet Spectra}

The {\fuse}\/ observatory consists of four aligned, prime-focus telescopes,
and Rowland-circle spectrographs that feed two photon-counting detectors.
Two of the telescope/spectrograph channels have SiC coatings to  
cover the range $\sim 905 - 1105$~{\AA}, while the other two have LiF 
coatings to cover the $\sim 980 - 1188$~{\AA} with high throughput.
These are referred to as the SiC and LiF channels, respectively. 
Each channel has its own focal plane assembly, which contains
three entrance apertures of different sizes.  After passing through 
one of these apertures, light is diffracted and focused by the concave 
gratings onto a delay-line detector, each of which records a pair of 
(SiC, LiF) spectra.  For faint objects, the detectors are operated in 
``time-tag'' mode to record the arrival time and detector position of 
individual photon events; for brighter sources, the incoming photons are 
recorded in ``histogram'' mode, which only retains positional information.
Since each detector is subdivided into two segments (labeled ``A'' and
``B''), eight independent spectra are ultimately obtained, one for each 
combination of channel and detector segment.  As a result, nearly the 
entire wavelength range between 905 and 1187~{\AA} is redundantly covered 
by at least two independent spectra.  Further details of the {\fuse}\/ 
mission, its instrumentation, and in-orbit performance are provided by 
\citet{Moos00} and \citet{Sahnow00}.

The observations presented here were obtained between 1999 December 15
and 2000 December 4, during the first year of the prime mission of {\fuse}.
All observations were made through the large, $30{\arcsec} \times 
30{\arcsec}$ (LWRS) aperture in time-tag mode, with typical integration 
times of $\sim$8~ks.  Although thermal motions of the telescope mirrors 
causes channel misalignment, the targets always remained in the LWRS 
apertures and no data were lost.

The spectra were processed uniformly with version 1.8.7 of the standard 
calibration pipeline software package, CALFUSE.  Processing steps included 
application of corrections for small, thermally-induced motions of the 
diffraction gratings; removal of detector backgrounds; correction for 
thermal and electronic distortions in the detectors; extraction of a 
one-dimensional spectrum by summing over the astigmatic height of the 
two-dimensional image; correction for the minor effects of detector dead 
time; and application of flux and wavelength calibrations. CALFUSE~1.8.7 
did not include corrections for residual astigmatism in the spectrograph 
or fixed-pattern noise in the detector, which limit the spectral 
resolution and S/N ratio of the extracted spectra, 
respectively.  Neither of these omissions are detrimental to this 
program, since analysis of wind profiles does not require extremely 
high spectral resolution, and since we always compared redundant data 
from different channels to determine the reality of weak spectral features.

However, the data were adversely affected by an anomaly known as ``the 
worm,'' which is a region of depressed flux caused by shadowing from grid
wires located near the detector to enhance its quantum efficiency 
{\citep{Sahnow00c}}.  The most severe worm affects the longest wavelength 
regions of spectra from the LiF1 channel recorded by detector segment B.
Fortunately, we were able to use the redundant information from LiF2
channel (detector segment A) to recover the true shape of the 
longest wavelength lines in our study ({\ion{P}{5}}~$\lambda\lambda$1118, 
1128).

Finally, the fully processed spectra, which have a nominal spectral 
resolution of $\sim$20~{\kms} (FWHM), were smoothed to a resolution 
of $\sim$30~{\kms} in order to enhance the S/N ratio.
Due to errors in the wavelength scale of CALFUSE 1.8.7, we found that  
systematic velocity shifts of about $-50$~{\kms} were required to place the 
strong interstellar H$_2$ lines due to absorption in the Milky Way at zero 
velocity.  The magnitude and sense of this shift agrees with the results 
of \citet{Danforth02} and \citet{Howk02}, who considered the problem in 
greater detail.

\subsection{Ancillary UV Spectra}

Fully processed, archival data sets of the targets obtained by {\iue}\/ 
or {\hst}\/ were retrieved from the Multimission Archive at Space Telescope 
(MAST).  The specific data sets are listed in Table~{\ref{datatab}}.
These spectra were subsequently smoothed in order to enhance their S/N.  
STIS spectra were smoothed to an effective resolution of 30~{\kms} in order 
to match the \fuse\/ data, while the high dispersion {\iue}\/ spectra 
(which were typically of quite poor quality at full resolution) were 
smoothed to 60~{\kms} to increase the S/N ratio. The FOS spectra were 
smoothed by 100~{\kms}, which is less than their intrinsic resolution of 
$\sim 240$~\kms.

\section{Wind Profile Analysis}

\subsection{Overview}

The shapes of P~Cygni wind profiles of resonance lines encode information 
about the velocity law governing the expansion of the wind; any additional 
macroscopic velocity fields that might be present, e.g., due to shocks; 
the abundance and distribution of the parent ion; and the total amount of 
material in the wind.  This information can be accessed by fitting the 
profiles subject to an underlying model for the structure of the wind.
Indeed, self-consistent fits are the only way to extract information
about the velocity law and ionic column densities as a function of
velocity in the wind.  This information is especially useful when several 
stars (or observations of the same star obtained over several epochs) are 
analyzed together, since any limitations of the underlying wind model are 
compensated to some extent by differential analysis of the trends 
exhibited by different ions.  Previous work based on {\iue}\/ spectra of 
O-type stars by \citet{Howarth89}, {\citeauthor{Groenewegen89b} 
\citeyearpar{Groenewegen89b,Groenewegen91}}, \citet{Haser95} and 
\citet{Lamers99} demonstrate the utility of this approach.

We employed a modified version of the ``Sobolev with Exact Integration" 
(SEI) computer program developed by \citet{Lamers87} together with 
a simplified model of the interstellar {\ion{H}{1}} and H$_2$ 
along the line of sight, to obtain precise fits to the P~Cygni profiles 
in the {\fuse} and {\iue}/{\hst} wavebands.  These fits yielded 
measurements of the run of wind optical depths as a function of velocity 
for the ion analyzed.  These were subsequently converted into the product 
of mass-loss rate and ionization fraction, {\mdotqi$(v)$}.  Finally, we 
used the {\citeauthor{Vink00} \citeyearpar{Vink00,Vink01}} formulae for 
mass-loss rates to convert the {\mdotqi$(v)$} measurements into 
velocity-dependent ionization fractions for each observed ion, $q_i(v)$. 
The following sub-sections describe the details of this procedure.

\subsection{The SEI Method}

The SEI method computes wind profiles for homogeneous, spherically
symmetric stellar winds characterized by smoothly accelerating
velocity laws.  It accounts fully for blending between the components of 
closely spaced doublets.  The inputs are specified in the following way.

\medskip
\noindent {\bf Terminal Velocity.}
A determination of {\vinf} is required to compute the normalized velocity 
parameter $w \equiv v(r) / ${\vinf}. 

\medskip
\noindent {\bf Velocity Law.} 
We assume a standard ``$\beta$-law'' for the expansion of the wind, which 
has the form
\begin{equation}
     w = w_0 + (1 - w_0) (1 - 1/x)^\beta , 
\end{equation}
where $x = r/R_\star$ and $R_\star$ is the stellar radius 
{\citep{Lamers87}}. We set $w_0 = 0.01$ in all cases.  The value of 
$\beta$ selected affects the overall shape of the profile, since it 
governs the density distribution, $\rho(x)$, through the equation of mass 
continuity:
\begin{equation}  
     {\dot M} = 4 \pi R_\star^2 v_\infty x^2 w( x ) \, \rho(x)~.
\end{equation}
Since $w$ is a monotonically increasing function of $x$ and the geometry
is assumed to be spherically symmetric, there is a one-to-one mapping 
between velocity and radial position in the wind.

The parameters of the velocity law are usually determined from a saturated 
wind line.  Profiles produced by small $\beta$ values ($\sim 0.5$) have 
shallower and broader red emission peaks than profiles with larger $\beta$ 
values.  The reason is that low $\beta$ winds accelerate more rapidly.  As 
a result, more of the low speed, red-shifted wind material is occulted by 
the stellar disk.  

\medskip
\noindent {\bf Turbulent Velocity.} 
The turbulent velocity field is characterized by a Gaussian distribution,
which is specified by the 1$\sigma$ dispersion parameter $w_D$.  This 
additional dispersion is intended to simulate the effects of macroscopic 
velocity fields due, e.g., to shocks in the wind, on the line profile.
It effectively smooths the distribution of optical depth as a function 
of $w$, which causes saturated portions of strong P~Cygni profiles to be
extended by a few times $w_D$.  Its primary effects are to decrease the 
sharpness of the absorption trough near {\vinf} and shift the maximum 
velocity seen in absorption blueward, and to shift the strength of the 
emission peak redward.  The latter effect is an artifact of the assumed 
constancy of $w_D$ throughout the wind, which usually implies that 
extremely large - and probably aphysical - velocity dispersions 
are present deep in the stellar wind; see, e.g., {\citeauthor{Haser95} 
\citeyearpar{Haser95,Haser98}} for discussion of this point.  In practice, 
use of a constant $w_D$ has little effect on subsequent analysis, since 
optical depths close to $w=0$ are excluded from consideration.

\medskip
\noindent {\bf Input Photospheric Spectrum.} 
In the absence of reliable templates for the wind-free FUV and UV flux
distributions of O-type stars, we assumed that the continuum was
either flat (i.e., no photospheric absorption lines corresponding
to the resonance lines under consideration) or that the photospheric
profiles were free parameters given by Gaussian distributions of optical 
depth 
\begin{equation}
   r_{w} = \exp \{ -\tau_0^B \exp [-w^2/\sigma_w^2]
                   -\tau_0^R \exp[-(w -\delta_w)^2/\sigma_w^2] \}, 
\end{equation}
where $\tau_0^R/\tau_0^B = f_R/f_B$, the ratio of oscillator strengths for 
the doublet (or zero for a singlet); $\delta_w$ is the spacing of the 
doublet in normalized velocity; and $\sigma_w$ is related to the full 
width of the line expressed as a velocity, $v_G$, by $v_G = 2\sqrt{\ln 2} 
\sigma_w v_{\infty}$.  Strengthening the photospheric spectrum affects the 
P~Cygni profile near $w=0$ by increasing the strength of the absorption 
trough and decreasing the strength of the emission lobe.

\medskip
\noindent {\bf Optical Depth of the Wind.} 
The optical depth of the wind is assumed to be given by the radial Sobolev 
optical depth, {\trad}$(w)$, which we modeled in a set of 21 independent 
velocity bins whose magnitudes were adjusted to obtain the best fit.  
This approach has been used previously by \citet{Massa95b}, and is similar 
to the technique described by \citet{Haser95}.  It has the advantage of 
avoiding biases due to preconceived notions about the functional form of 
the distribution of {\trad}$(w)$.  It also simplifies the fitting 
procedure, since {\trad}$(w)$ is the only free parameter once the 
velocity law and photospheric spectrum are fixed.  This fitting scheme 
relies on the fact that in a monotonically expanding, spherically 
symmetric outflow, only material with $w \le w_i + w_D$ contributes to 
the formation of the line profile at $w_i$; see, e.g., the discussion of 
P~Cygni profile formation in \citet{LCbook99}.  Therefore, the profiles 
can be fit by first determining the value of {\trad}$(w)$ that fits the 
profile at $w \approx 1$, and then stepping inward through the absorption 
trough {\citep[see][]{Massa95b}}.  At each step $i$, the value of 
{\trad}$(w_i)$ is changed until a satisfactory fit to the profile at 
$w_i$ is achieved.  As long as the fundamental assumptions of the model 
are correct, particularly a monotonicity of the velocity law and spherical 
symmetry, reliable values of {\trad}$(w)$ are derived.

\medskip
\noindent{\bf Ion Fractions.}
While {\trad} determines the shape of the profile, the more physically 
meaningful quantity is the ionization fraction, $q_i(w)$.  The ionization 
fraction is related to {\trad} by {\citep[see, e.g.,][]{Olson82}}
\begin{equation}
  q_i(w) = \left( \frac{m_e c}{\pi e^2}\right) \frac{4 \pi \mu m_H}
 {f \lambda_0 A_E} \frac{R_\star v_{\infty}^2}{\dot{M}} x^2 w 
 \frac{dw}{dx} \tau_{rad}(w) \label{qi}
\end{equation}
where $q_i(w)$ is the fraction of element $E$ in ionization state $i$ at 
velocity $w$; {\mdot} is the mass-loss rate of the star; $\mu$ is the 
mean molecular weight of the plasma (which was set to 1.35 for all 
program stars); $A_E$ is the abundance of element $E$ relative to 
hydrogen by number and all other symbols have their usual meaning.  
Adopted values of the oscillator strengths and elemental abundances are 
given in Table~{\ref{atomtab}}.  

\subsection{Procedure to Fit Wind Profiles}

{\fuse} spectra are affected by a wealth of interstellar absorption lines,
which include numerous H$_2$ lines, the upper Lyman series of {\ion{H}{1}},
and several strong lines from metallic ions such as {\ion{O}{1}}, 
{\ion{N}{1}}, {\ion{C}{2}}, and {\ion{Si}{2}} 
{\citep[see, e.g.,][]{Friedman00}}.  As a result, the FUV spectrum of even 
a lightly reddened star is strongly affected by ISM lines which complicates 
the analysis considerably {\citep[see][]{Bianchi00,Fullerton00}}.  This 
plethora of strong interstellar lines throughout the {\fuse}\/ range 
affects the analysis in two respects.  First, a rough model of the ISM 
absorption is required in to disentangle the effects of stellar and 
interstellar absorption.  Second, the resulting complexity of the spectrum 
makes any automated  fitting scheme effectively impossible.  Consequently, 
all fits were performed interactively.

The first step of the analysis for each star was to estimate interstellar
{\ion{H}{1}} and H$_2$ column densities from interactive fits of a simple 
model to the observed interstellar lines.  The model consisted of two 
{\ion{H}{1}} components, one appropriate to the Galaxy and one appropriate 
to the LMC, and a single component for Galactic H$_2$.  For the H$_2$ 
model, we fit each rotational level with its own column density and 
velocity spread parameter (i.e., $b$-value).  This resulted in cosmetic 
fits that were suitable for our purposes, though the derived column 
densities are not expected to be very reliable.  To determine accurate 
column densities, a detailed kinematic model for the ISM along the line 
of sight would be required.

Next, whenever possible, we used a saturated wind line to determine the 
parameters of the velocity law: $\beta$, {\vinf}, and $w_D$.  We typically 
used the {\ion{C}{4}} $\lambda\lambda$1548, 1550 doublet for this purpose, 
since it is usually strongly saturated and hence primarily sensitive to 
the velocity field \citep[see, e.g.,][]{Kudritzki00}.  However, since the 
{\iue}\/ or {\hst}\/ spectra containing the {\ion{C}{4}} feature were 
obtained at different epochs than the FUV spectra, and since the ``blue 
edges'' of P~Cygni absorption troughs are often variable 
{\citep[see, e.g.,~][]{Kaper96}}, small but significant differences in 
{\vinf} were permitted between the data sets.  We also allowed $w_D$ to 
differ for each line, since different ions sample different components of 
the wind plasma. 

Once {\vinf} and $\beta$ were determined, all the resonance lines were 
fit by eye by adjusting the histogram representation of {\trad} in the 
manner described in \S 3.2 and the parameters describing the photospheric
spectrum, $\tau_o^B$ and $v_G$.  The process is summarized in 
Figure~\ref{worksheet}, which illustrates the final results of the 
iterative procedure to fit {\ion{S}{4}} in {\sk{-67}{111}} (left) and 
{\ion{O}{6}} in {\sk{-70}{69}} (right).  There are three panels for each 
line; in all cases, the velocity scale is in the stellar rest frame, and 
refers to the blue component of a doublet.
\begin{enumerate}
  \item The top panel shows the computed stellar wind profile overplotted 
        on the observed spectrum.  The rest position of the red component 
        of the doublet is indicated by a thick tick mark along the 
        (normalized) velocity axis.  The model profile is also decomposed 
        into its direct (transmitted) and diffuse (scattered) components, 
        which are shown as dotted and dashed lines, respectively.  This 
        decomposition provides useful feedback during the ``outside--in'' 
        fitting procedure, because the addition of optical depth to 
        decrease the transmitted flux at some high velocity will 
        necessarily increase the forward-scattered component at lower 
        velocities in the absorption trough of the P~Cygni profile.  Once 
        absorption at high velocity produces more scattered light than 
        the observed profile at lower velocity (in the same component of 
        the doublet), then even the addition of arbitrarily large optical 
        depths at low velocities will not deepen the profile there.  In 
        this case, the only way to increase the low-velocity absorption 
        in the model profile is by increasing the strength or breadth of 
        the adopted photospheric profiles.  If that fails to improve the 
        fit, then the validity of the model assumptions -- especially 
        spherical symmetry -- must be questioned.
        
  \item The middle panels show the derived value of {\trad} in the
        21 velocity bins (left) and the input photospheric spectrum
        (right), both in units of $w$.
        
  \item The bottom panel compares the observed profile with the best-fit
        model, which has been convolved with the adopted spectral
        smoothing (\S 2.2, 2.3).  Whenever appropriate, the best fit model 
        includes the effects of {\ion{H}{1}}~+~H$_2$ absorption from the 
        ISM. 
        
\end{enumerate}
                
Finally, once the entire suite of wind profiles for a given star had been
fit, we performed a final consistency check, which often resulted in the 
parameters of one or more lines being adjusted to provide better internal 
agreement.

\section{Resonance Line Diagnostics}

\begin{figure}
\figurenum{1}
\epsscale{0.90}
\plotone{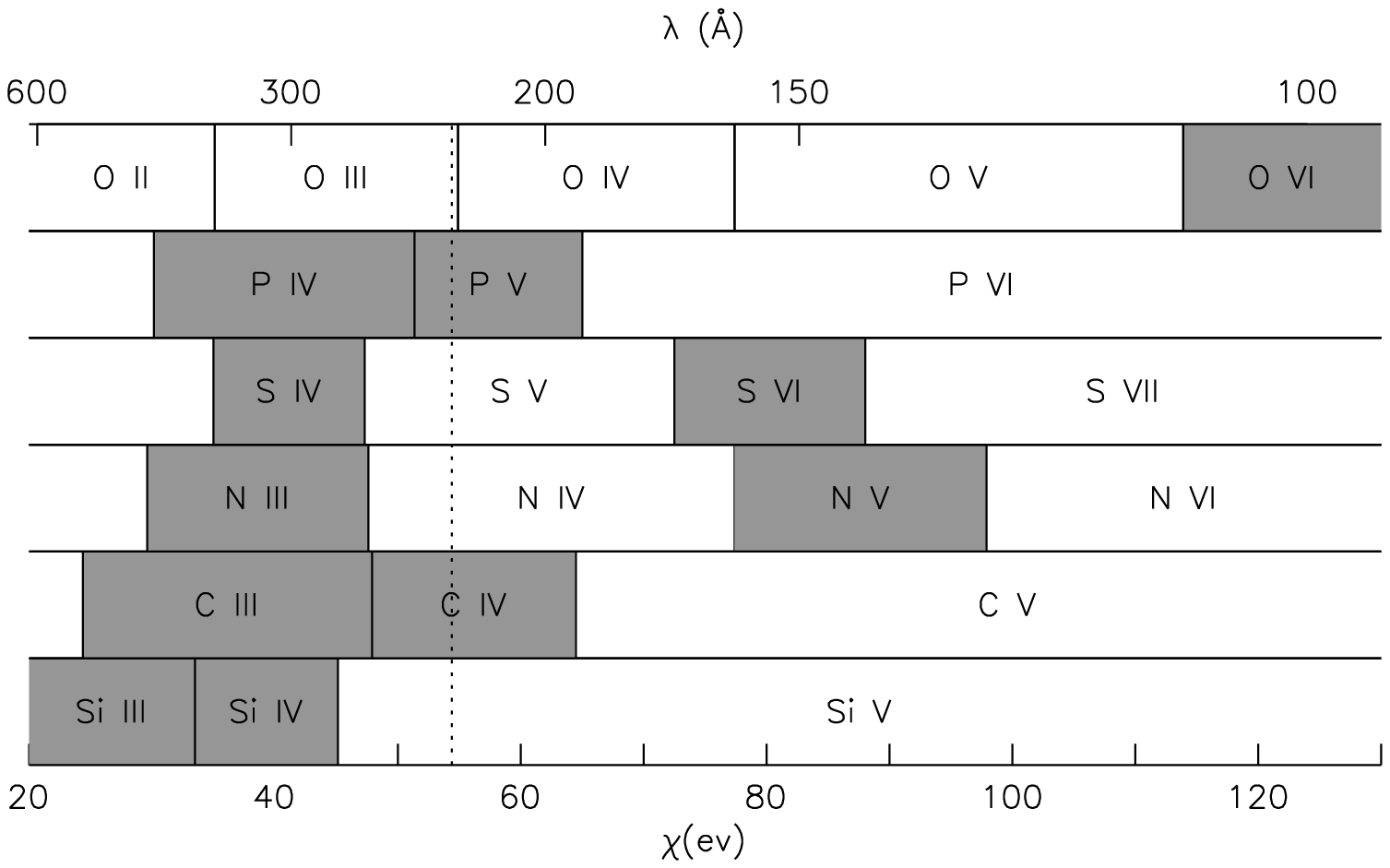}
\caption{
  Schematic representation of the range in ionization spanned by ions that 
  produce stellar wind lines in spectra of O-type stars.  Ions in shaded 
  boxes are those with observed resonance lines, while ions in unshaded 
  boxes do not have resonance lines in the FUV or UV.  Open-ended boxes 
  indicate that the ion persists beyond the range of the plot.  The dashed 
  vertical line indicates the ionization potential of {\ion{He}{2}}, 
  54.4~eV.  Ions that lie completely to the right of this line are 
  ``super-ions,'' i.e., ions that cannot be abundantly produced by 
  photoionization from the ground state, since the stellar continuum flux 
  shortward of 228~{\AA} is drastically reduced by bound-free absorption 
  from {\ion{He}{2}}.
  \label{ions}
}
\end{figure}

\begin{figure}
\figurenum{2}
\epsscale{1}
\plottwo{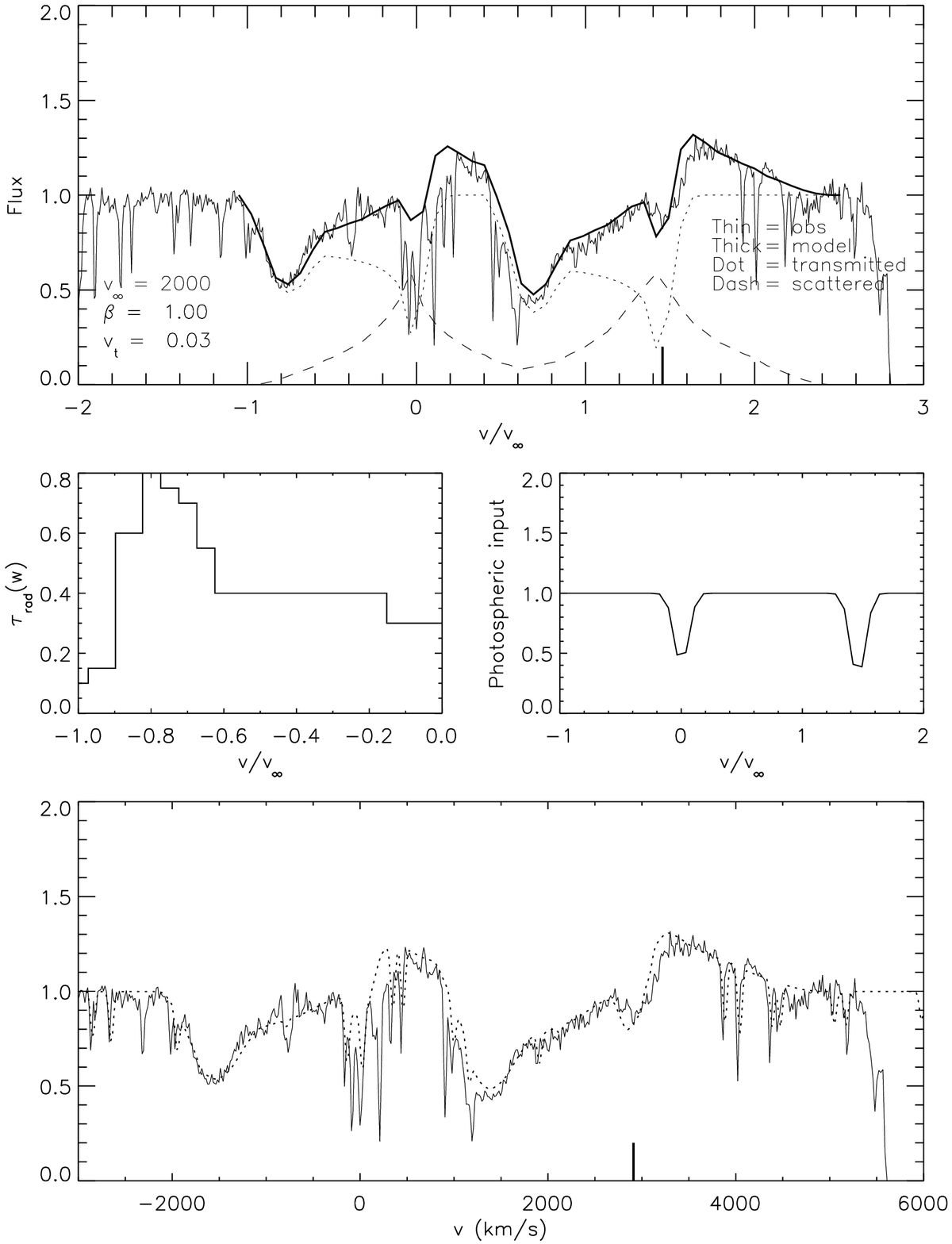}{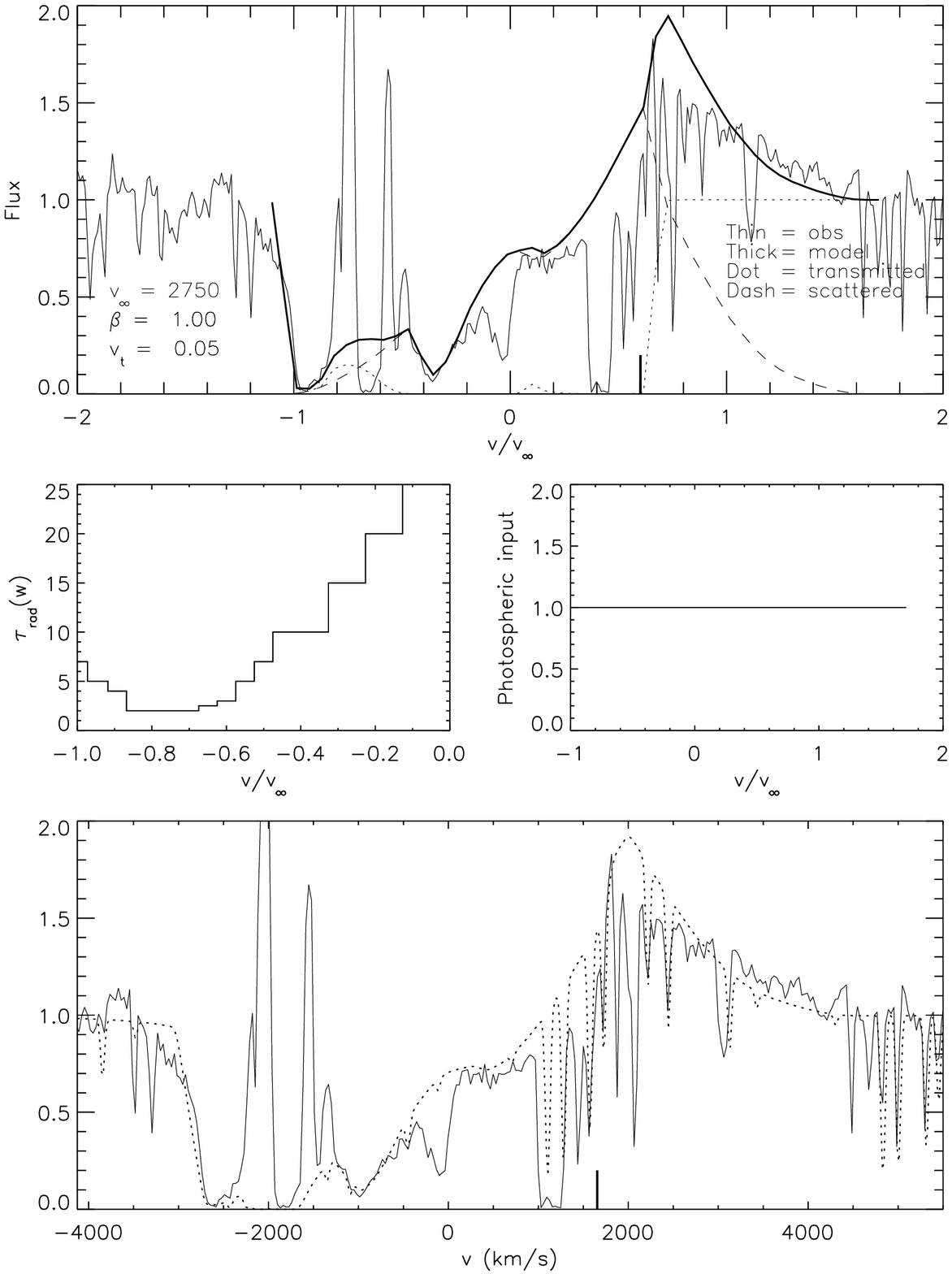}
\caption{Examples of fits for {\ion{S}{4}} in {\sk{-67}{111}} (left) and
  {\ion{O}{6}} in {\sk{-70}{69}} (right).  The velocity scales are in the 
  stellar rest frames.  The top panels show the raw model output, the 
  observed spectrum, and the contributions to the model profile from 
  transmitted and scattered light as functions of normalized velocity.  
  A thick tick mark denotes the rest position of the red component of the 
  doublet.  The middle panels show {\trad} and the input photospheric 
  profile as a function of normalized velocity.  The bottom panels compare 
  the observed profile to the model after application of an {\ion{H}{1}} 
  $+$ H$_2$ ISM model and convolution with the instrumental profile. 
  See \S 3.3.
  \label{worksheet}
}
\end{figure}

The complete suite of resonance line diagnostics obtained by combining 
spectra from {\fuse}\/ and {\iue}\/ or {\hst}\/ is illustrated 
schematically in Figure~\ref{ions}, which shows the energy required to 
ionize each species; see also Table~\ref{atomtab}.  The ions uniquely 
accessible to {\fuse}\/ in Fig.~\ref{ions} provide unmatched information on 
the ionization structure of the wind by bracketing the species accessible 
to {\iue}\/ or {\hst}\/ and extending the range of energies that can be 
probed.  The non-CNO ions of sulfur and phosphorus are particularly useful 
since the abundances of these elements, like silicon which is normally 
studied at longer wavelengths, are unaffected by the nuclear processes that 
occur throughout the hydrogen burning lifetime of a massive star.  
Furthermore, because they have much smaller cosmic abundances, the 
resonance lines of these ions are less likely to be saturated, even when 
they are near the dominant stage of ionization.  Hence, they are valuable 
probes even for very dense winds.

Specific comments about the importance of each of these ions follow along 
with particular aspects of the spectra that affected our fitting 
procedures.  The discussion is arranged in order of decreasing ionization
potential.

\medskip
\noindent{\bf {\ion{O}{6}}~$\lambda\lambda$1031, 1037.}  
The super-ion {\ion{O}{6}} is the best diagnostic of high-energy processes 
in the optical/UV region of the spectrum, and provides a direct link with 
the distribution and strength of X-rays in the wind.  Unlike {\ion{N}{5}} 
and {\ion{S}{6}}, the next lower ion -- {\ion{O}{5}} -- is also a super-ion.
The fact that two electrons must be removed from {\ion{O}{4}}, the dominant 
stage of O, in order to produce the observed {\ion{O}{6}} lines was the 
clue that lead \citet{Cassinelli79} to suggest that X-rays are responsible 
for the production of {\ion{O}{6}} via Auger ionization. 

Fits to the {\ion{O}{6}} doublet are affected by blends with the 
interstellar Ly~$\beta$ line, which lies 1805~{\kms} blueward of the 
{\ion{O}{6}} $\lambda$1031 component.  Since Ly~$\beta$ is invariably 
saturated, it is difficult to determine {\trad} in the wind profile beyond 
this velocity.  However, the wide separation of the doublet (1654~{\kms}) 
permitted the red component to be used to constrain {\trad} at high 
velocity.  Additional constraints were possible in the few cases where 
{\vinf} was large enough to emerge on the blue side of the Ly~$\beta$ 
absorption.  In addition to Ly~$\beta$, there is often sizable ISM 
absorption due to {\ion{O}{6}} itself near line center, strong {\ion{C}{2}} 
lines near the rest wavelength of the red component, numerous {\ion{O}{1}} 
lines in the absorption trough and several strong H$_2$ lines.  Significant 
emission due to airglow in Ly~$\beta$ and the {\ion{O}{1}} multiplet 
between 1027 and 1029~{\AA} accumulates in long exposures.  We did not 
include the effect of photospheric lines of {\ion{H}{1}} or {\ion{He}{2}} 
in the fits.

\medskip
\noindent {\bf {\ion{N}{5}}~$\lambda\lambda$1238, 1242.} Although 
{\ion{N}{5}} is a super-ion, it is only one stage above the dominant ion 
for most O stars.  Owing to the large abundance of nitrogen, the resonance 
lines are frequently saturated, and hence of limited diagnostic utility.  
The broad interstellar Ly~$\alpha$ absorption affects the high velocity 
portion of the wind line and was included in the fits.

\medskip
\noindent{\bf {\ion{S}{6}}~$\lambda\lambda$933, 944.} Like {\ion{N}{5}}, 
{\ion{S}{6}} is a super-ion that is just one stage above the dominant ion 
for the hotter O stars.  However, unlike {\ion{N}{5}}, {\ion{S}{6}} is 
rarely saturated since the sulfur abundance is only $\sim$16\% that of 
nitrogen (Table~\ref{atomtab}).  Well developed 
{\ion{S}{6}}~$\lambda\lambda$~933,~944 absorption was present in all of 
the program stars except {\sk{-68}{135}} (ON9.7~Ia+).

This spectral region is by far the most difficult to model because it is 
strongly affected by the confluence of both stellar and interstellar 
{\ion{H}{1}} Lyman lines and the stellar {\ion{He}{2}} Balmer series.  The 
region also suffers absorption by interstellar H$_2$, and extinction by 
dust.  Depending on the terminal velocity of the wind and the strengths of 
{\ion{P}{4}}~$\lambda$950 and {\ion{N}{4}}~$\lambda$~955, wind lines from 
these two ions can impinge on the red component of the {\ion{S}{6}}.  
Furthermore, the blue component may be affected by blending with the 
excited {\ion{N}{4}} $\lambda$923 feature in very dense winds.  All of 
these factors make it extremely difficult to define the stellar continuum 
in this region; see also \citet{Bianchi00}.  Since we do not have accurate 
models to account for the effects of any of the stellar blends, we simply 
ignored them.

\medskip
\noindent{\bf {\ion{P}{5}}~$\lambda\lambda$1117, 1128.}  Although 
{\ion{P}{5}} spans a range of ionization energy that is only slightly 
higher than {\ion{C}{4}}, its resonance doublet is never saturated because 
it is a thousand times less abundant than carbon (Table~\ref{atomtab}).
Unfortunately, this low abundance also makes 
{\ion{P}{5}}~$\lambda\lambda$1117, 1128 detectable only in stars with very 
massive winds.  Nevertheless, these lines are of great importance, since 
{\ion{P}{5}} is expected to be near the dominant stage of ionization.

Although there is little interstellar contamination in the vicinity of the 
{\ion{P}{5}} lines, they are often blended with the 
{\ion{Si}{4}}~$\lambda\lambda$~1122, 1128 lines, which arise from an 
excited state.  It is extremely difficult to disentangle the effects of 
this blend.  Consequently, we gave the {\ion{P}{5}}~$\lambda$1117 
component of the doublet higher weight in our fitting.  However, it is 
possible that the some {\ion{P}{5}} column densities are overestimated 
because contributions from the {\ion{Si}{4}} were neglected.

\medskip
\noindent{\bf {\ion{C}{4}}~$\lambda\lambda$1548, 1550.}  Since it is an 
intrinsically abundant species near the dominant stage of ionization, the 
resonance lines of {\ion{C}{4}} are generally saturated in spectra of LMC 
O stars.  Although they provide only lower limits on {\trad}, they are 
excellent diagnostics of the parameters of the velocity law.  We normally 
use this doublet to determine $\beta$ and {\vinf}.

The only ISM lines affecting the {\ion{C}{4}} fits are from interstellar 
{\ion{C}{4}} in the Galactic halo and the LMC.  However, the general region 
can be affected by photospheric line blanketing in cooler, more luminous O 
stars.

\medskip
\noindent{\bf {\ion{P}{4}}~$\lambda$950.} {\ion{P}{4}} lies one stage 
below the expected dominant stage of ionization in the winds of early 
O-type stars.  Since the intrinsic abundance of phosphorus is very low 
(Table~\ref{atomtab}), sufficient column densities for secure detection 
occur only for late O-type stars with very massive winds.  However, even 
when it is detectable, the line is blended with interstellar Ly~$\delta$ 
and the nearby {\ion{S}{6}} doublet, which makes reliable measurements of 
the profile extremely difficult.  Consequently, no quantitative 
measurements are given for the {\ion{P}{4}} resonance line.

\medskip
\noindent{\bf {\ion{C}{3}}~$\lambda$977.}  The large oscillator strength 
of this transition, together with the high abundance of carbon, allow it 
to persist (and even be quite strong) into the early O stars, even though 
it is two stages below the dominant ion.  Unfortunately, since the 
{\ion{C}{4}} resonance lines are almost always saturated in the winds of 
early O stars, the combination of {\ion{C}{3}} and {\ion{C}{4}} is not 
very useful.  

Fits to {\ion{C}{3}}~$\lambda$977 wind profiles are compromised by blends 
with several strong ISM lines, including {\ion{C}{3}} from both the 
Galactic halo and LMC, H$_2$,  {\ion{O}{1}} near the rest velocity, and 
{\ion{N}{1}} and the saturated Ly~$\gamma$ line in the P~Cygni absorption 
trough.  Furthermore, since this line is less than 4000~{\kms} from 
{\ion{N}{3}}~$\lambda\lambda$990, 991, its emission lobe will be affected 
by blueshifted {\ion{N}{3}} absorption  for stars with 
{\vinf}~$\gtrsim$~2000~{\kms}.  

\medskip
\noindent{\bf {\ion{N}{3}}~$\lambda\lambda$990, 992.} {\ion{N}{3}} is 
expected to be one stage below the dominant ion.  Together with 
{\ion{N}{5}}, it forms a pair similar to {\ion{S}{4}} and {\ion{S}{6}}.  
However, it is rare that both the {\ion{N}{3}} multiplet and the 
{\ion{N}{5}} doublet are present and unsaturated.  In our sample, this 
only occurs for stars of intermediate luminosity near spectral class O8.

The {\ion{N}{3}} resonance transition is a multiplet of three components,
with wavelengths of 989.799, 991.511 and 991.577~{\AA} 
(Table~\ref{atomtab}).  However, since the 991.511~{\AA} component is 
roughly ten times weaker than the other two components, and because our 
SEI program can only accommodate two overlapping components, we combined 
the oscillator strengths of 991.511 and 991.577\AA\ and represent them as 
a single line; i.e., we represent the multiplet as a doublet.  

As with {\ion{C}{3}}~$\lambda$977, the {\ion{N}{3}} resonance line is 
strongly affected by ISM absorption.  In this case, the problems arise 
from {\ion{N}{3}} itself and {\ion{Si}{2}} near line center, {\ion{O}{1}} 
to the blue and H$_2$ throughout the region.  A further complication is the 
presence of a group of {\ion{O}{1}} airglow lines between 988 and 991~{\AA} 
that are quite strong, particularly in long exposures on faint objects.

\medskip
\noindent{\bf {\ion{S}{4}}~$\lambda\lambda$1062, 1073.}  {\ion{S}{4}} is 
one stage below the expected dominant ion in the winds of early O stars. 
It exists over essentially the same range of ionization energies as 
{\ion{Si}{4}}, though sulfur is only half as abundant as silicon 
(Table~\ref{atomtab}).  The simultaneous presence of {\ion{S}{4}} and 
{\ion{S}{6}} in dense winds provides a rare opportunity to study the 
behavior of unsaturated resonance lines for two ions of the same species.
Although neither species is dominant, ratios of their optical depths are 
independent of both the sulfur abundance and the stellar mass-loss rate.

The blue component of the {\ion{S}{4}} doublet is a resonance line, while 
the red component actually consists of two closely spaced fine structure 
lines with wavelengths of 1072.974 and 1073.516~{\AA}.  They arise from a 
level that lies 0.12 ~eV above ground.  Since the 1072.974~{\AA}\ line is 
9 times stronger than 1073.516~{\AA} line, we represented them as a single 
transition.  Furthermore, since the fine-structure lines lie so close to 
the ground state, we treated the 1062 and 1073~{\AA} lines as a resonance 
doublet.  During the analysis, it became apparent that either the ratio of 
the oscillator strengths of the \ion{S}{4} lines is incorrect, or there is 
an additional, unknown, line affecting the red component.  The only way we 
could obtain good fits for the red component was to increase its oscillator 
strength by 33 -- 50\%.  We adopted the more conservative value of 33\%.  
Since all quantitative results are based on optical depths determined from 
the blue component, the effect of increasing the strength of the red 
component is essentially cosmetic.  Nevertheless, it does indicate a 
problem with for these lines.   Although there is minimal contamination 
from interstellar absorption in this region, we often had to include 
photospheric lines in the fit.

\medskip
\noindent {\bf {\ion{Si}{4}}~$\lambda\lambda$ 1393, 1402.} {\ion{Si}{4}} 
is a trace ion in the winds of most O stars.  Its resonance lines are the 
only wind features that are generally unsaturated in the wavelength region 
accessible to {\iue}\/ and {\hst}.

Although interstellar absorption features from {\ion{Si}{4}} in the 
Galactic halo and the LMC are present, they have minimal effect on fits to 
the wind lines.  We did not account for the photospheric line blanketing 
that is particularly evident in spectra of cooler, more luminous O stars;
see, e.g., \citet{Haser98}.

\section{Results}

\begin{figure}
\figurenum{3}
\epsscale{0.90}
\plotone{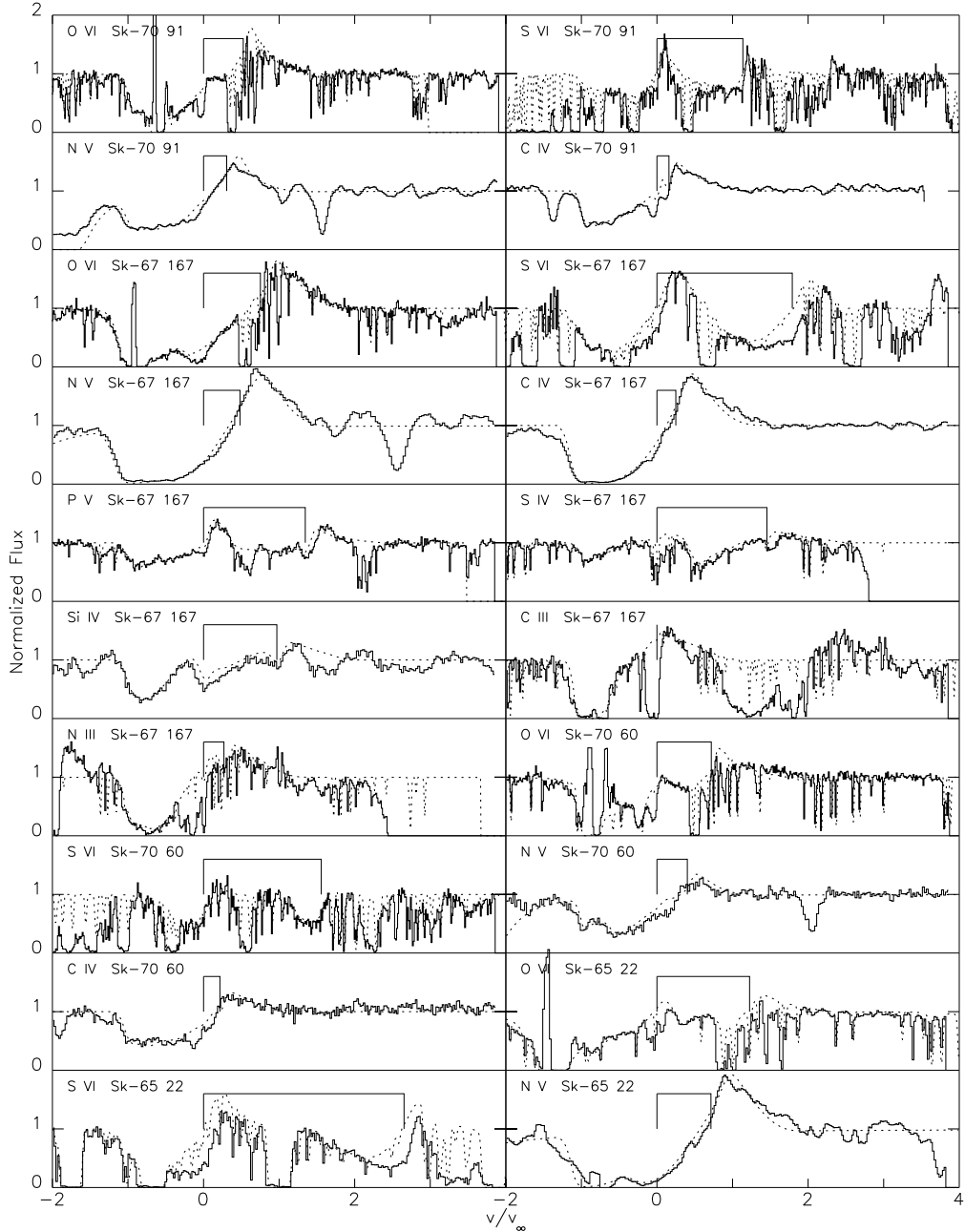}
\caption{
   Observed wind profiles (solid curves) and fits (dotted curves) for 
   selected program stars as a function of normalized velocity.  The fits 
   incorporate crude interstellar absorption models to demonstrate the 
   impact of ISM absorption on the profiles.  The name of the ion, star, 
   and the rest position of the line or doublet are indicated in each 
   panel.
   \label{fits1}
}
\end{figure}

\begin{figure}
\figurenum{4}
\epsscale{0.90}
\plotone{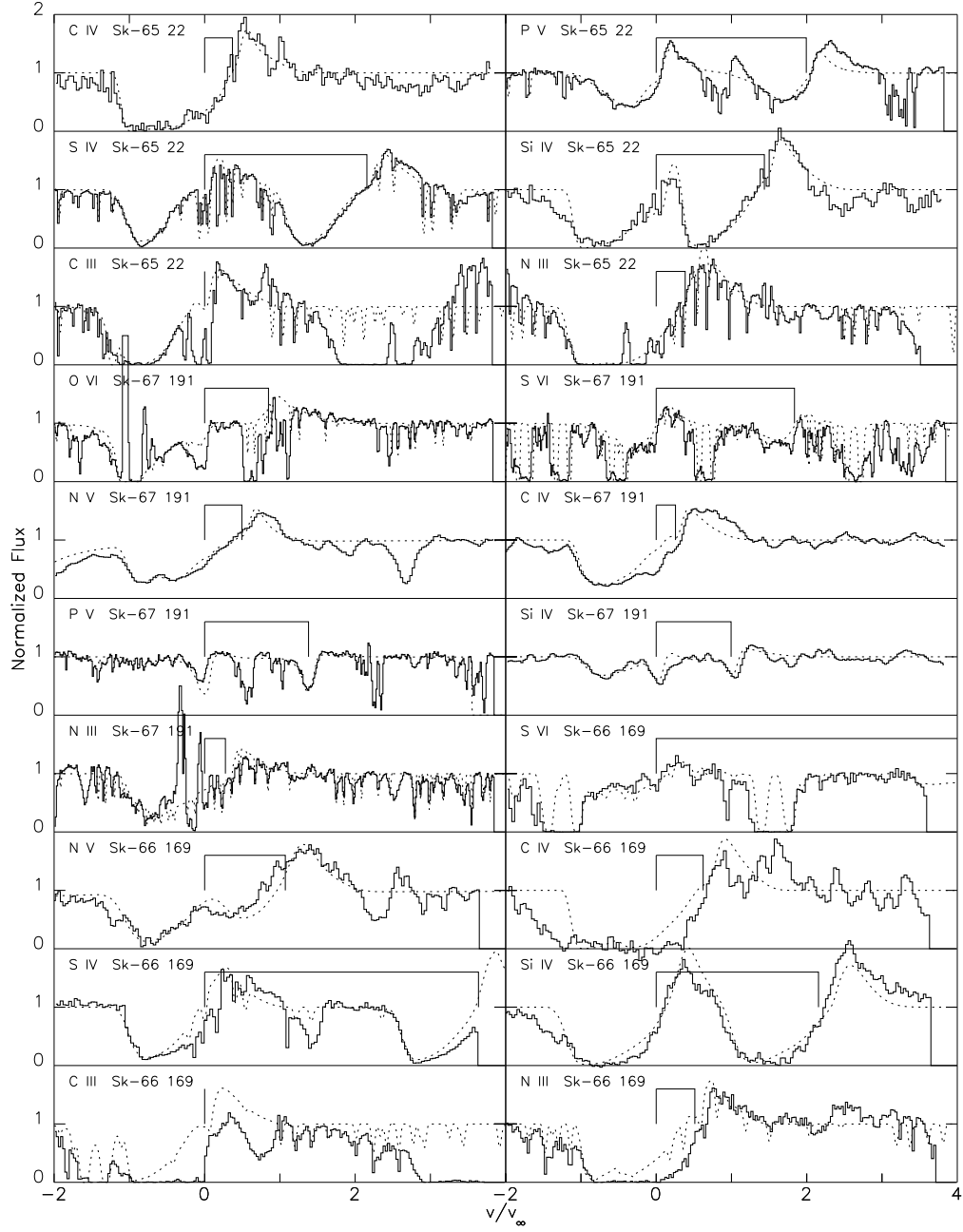}
\caption{Same as Figure \ref{fits1}
   \label{fits2}
}
\end{figure}

Figures~\ref{fits1} and \ref{fits2} show examples of the final fits to 
the available wind lines for several stars.  The normalized, observed 
profiles are shown as solid lines as 
a function of $w$, while the fits are shown as dotted curves.  Whenever 
appropriate, the fits incorporate a crude model spectrum of interstellar 
{\ion{H}{1}} and H$_2$ absorptions. The name of the star and the identity 
of the ion are given in the upper left-hand corner of each panel, and the 
rest positions of the components of the doublet are indicated above the 
spectrum.

For each star, the results of the fitting consist of (a) the parameters of 
the velocity law, {\vinf} and $\beta$; and, for each ion analyzed: (b) the 
parameters of the input photospheric profile, $\tau_0^B$ and $v_G$; (c) 
the turbulent velocity parameter, $w_D$; and (d) a tabulated set of 21 
$\tau_{rad}(w)$ values.  Each of these parameters and their associated 
errors will be discussed in the following subsections.  First, however, 
we discuss a few systematic effects in the modeling procedure that might 
affect the derived quantities.  The first is a trade-off between the 
presence of structure in {\trad}$(w)$ and the functional form of the 
velocity law.  This can be seen in Equation~[6], which shows that 
{\trad}$(w)$ is directly proportional to {\mdotqi}$(w)$ and inversely 
proportional to the velocity gradient.  Therefore, it is possible that 
some of the structure seen in the derived distribution of {\trad}$(w)$ 
(see, e.g., Figure~\ref{worksheet}) is caused by a mismatch between 
the real velocity law and its assumed functional form.  In particular, 
the cusps in {\trad} that often occur near $w=1$ might be an artifact of 
the actual velocity law having a flatter slope at high $w$ than the adopted 
$\beta$ law, which can only be compensated by a localized increase in 
the optical depth.  If only a single ion shows a localized enhancement, 
then it is likely there is a peculiarity in the formation of the ion.
Alternatively, if all the ions show the same cusp, then we conclude
that either there is a poor match between the actual velocity law and
the standard $\beta$ law, or there is a genuine enhancement in density
at that velocity due, e.g., to time-dependent changes in the structure
of the wind.

Second, because we ignore photospheric Ly $\beta$ and the upper 
{\ion{He}{2}} Balmer lines when we fit {\ion{O}{6}} and {\ion{S}{6}}, 
the models may produce too much emission for the observed absorption.  
This is because the model photosphere may have more photospheric flux 
available to be scattered by the wind than is present in the actual 
photosphere.  

Finally, since the model used to fit the profiles contains certain 
idealizations (e.g., it is spherically symmetric, homogeneous and steady), 
it is possible that some stellar profiles cannot be fit because the
structure of their winds is in fact more complicated.

In view of these complications, it is difficult to assign uncertainties
in the derived quantities rigorously.  This difficulty is exacerbated by 
several selection effects that also influence the quality of the fits.
These include: the strength of the line, since lines with optical depths 
near unity are most sensitive to changes in the parameters; the intrinsic 
spacing of the doublet relative to {\vinf}, which determines the extent 
to which the components are blended; whether there is overlap with wind 
lines from other ions; and the strength of the ISM absorption in the 
vicinity of a particular line.  The estimated uncertainties in the 
derived parameters given below represent a subjective assessment of 
the errors based on experience.  With few exceptions, changing the input 
parameters by the stated amounts would result in a significant change in 
the computed line profile, and a worse fit to the observed spectra.

\subsection{Photospheric Parameters}

Table~{\ref{phototab}} lists the photospheric data derived from the fits.  
The first column gives the star name, and the remaining columns give the 
values of $\tau_0^B$ and $v_G$ (see Equation [5]) for each line fit.  
Ellipsis indicate that the line was not strong enough to fit; entries 
of $0$ imply that the input photospheric spectrum was flat.  The 
uncertainties in both quantities could be as large as $\sim$50\%.

The stellar continua underlying the {\ion{O}{6}} $\lambda\lambda$1032, 
1038 lines were generally assumed to be flat, and hence are not included 
in Table~{\ref{phototab}}.  The only exception was {\sk{-67}{167}}, for 
which optimal fits required ($\tau_0^B$, $v_G$ ) = (4.9, 200~{\kms}).

\subsection{Velocity Parameters}

Table~{\ref{veltab}} summarizes the derived parameters associated with the 
velocity field of the wind.  Successive columns list the star name, 
{\vinf} (which is also given in Table~{\ref{startab}}), $\beta$, and the 
derived values of $w_D$ for each ion.  The uncertainty in {\vinf} is 
$\sim \pm$150~{\kms} (determined below), while $\beta$ is expected to 
be accurate to $\pm$0.25.  The uncertainties in the values of $w_D$ are 
$\sim \pm$0.05. 

Values of $\beta$ vary from 0.5 (``fast'') for the main sequence star 
{\sk{-70}{60}}  [O5-6 Vn((f))] to 2.0 (``slow'') for the late O supergiant 
{\sk{-66}{169}} [O9.7 Ia$+$], which also has the lowest {\vinf} in the 
sample.  The values of $w_D$ do not vary significantly from one line to 
another, though values derived from the lower resolution FOS data tend to 
be a bit larger.

\begin{figure}
\figurenum{5}
\epsscale{0.80}
\plotone{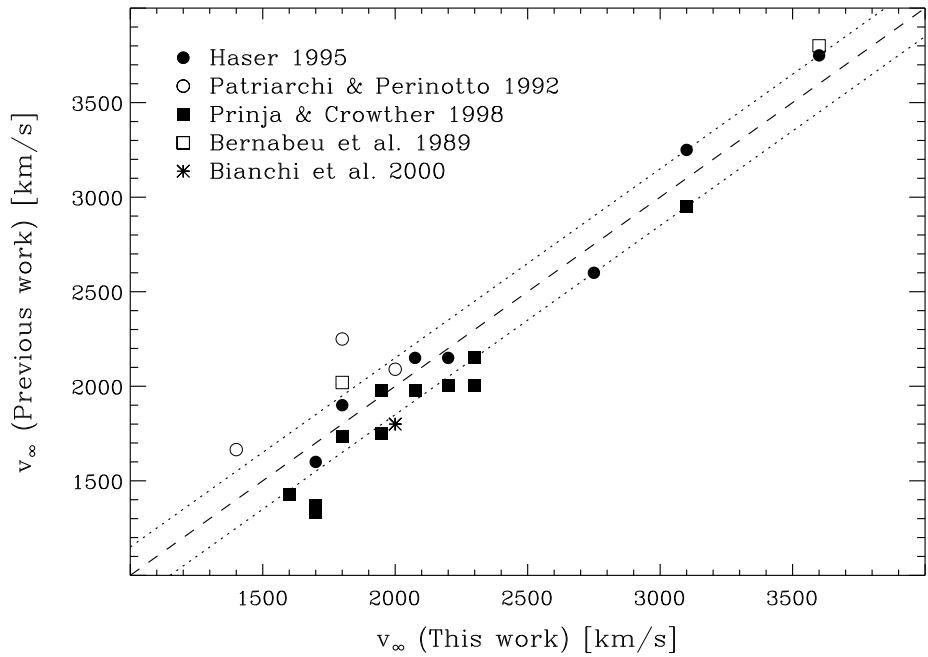}   
\figcaption{
   Comparison between between previous determinations of {\vinf} and the
   values derived in the present work.
   The dashed line indicates a one-to-one correlation, while the
   dotted lines denote offsets of $\pm$150~{\kms} from this line.
   \label{cfvinf}
   }
\end{figure}

Our determinations of {\vinf} and {$\beta$} are compared with previous 
measurements in Table~\ref{cfvlaw}.  The values of {$\beta$} are likely 
influenced by a variety of subtle biases, but are in good agreement within 
the adopted uncertainty.  The values of {\vinf} are also compared in 
Figure~\ref{cfvinf}, which shows agreement to within $\sim$150~{\kms} 
(i.e., to better than 10\%).  Previous determinations based on the 
positions of diagnostics of wind structure by \citet{Prinja98} tend to 
underestimate {\vinf}, perhaps because these direct measurements were made 
from comparatively low-resolution spectra obtained with FOS.  As expected, 
measurements of the maximum velocity seen in absorption by 
\citet{Bernabeu89} overestimate {\vinf}, though the measured values are 
in good agreement if they are assumed to be extended by $\sim$10\% due 
to the presence of macroscopic ``turbulent'' velocity fields.  Especially 
good agreement is found with the profile fits performed by \citet{Haser95}.

\subsection{Radial Optical Depths}

The uncertainties in measured values of {\trad} are subject to a variety
of selection effects, but are estimated to be:
  $\pm$50\% for a single velocity bin for {\ion{N}{3}}, {\ion{C}{3}}, 
            and {\ion{S}{6}};
  $\pm$50\% for {\ion{N}{5}}, {\ion{C}{4}}, and {\ion{Si}{4}} when derived 
            from FOS data;
  $\pm$30\% for {\ion{N}{5}}, {\ion{C}{4}}, and {\ion{Si}{4}} when derived 
            from high resolution {\iue}\/ or STIS data;
 and 
  $\pm$20\% for {\ion{S}{4}}, {\ion{P}{5}} and {\ion{O}{6}}. 

Saturation is an additional complication associated with measurements
of {\trad}.  It occurs when a model profile no longer responds to changes 
in the optical depth at a level that significantly affects the match with 
the observed profile.  For our data, this tends to occur at {\trad}$(w) 
\sim 3$ for singlets, doublets with similar oscillator strengths (i.e., 
{\ion{S}{4}}), and closely spaced doublets (i.e., {\ion{N}{3}}, 
{\ion{C}{3}}, {\ion{C}{4}}, and {\ion{N}{5}}).  Additional leverage is 
possible in the case of widely spaced doublets (i.e., {\ion{Si}{4}}, 
{\ion{P}{5}}, {\ion{S}{6}}, and {\ion{O}{6}}), since the weaker, red 
component provides information until its {\trad} $\sim$3, which 
corresponds to {\trad} $\sim$6 in the blue component.  Thus, in the case 
of saturated velocity bins, we adopted a lower limit of either 3 or 6 for 
{\trad} of the blue component, depending on the separation of the doublet.

Upper limits are also difficult to assign rigorously, since {\trad}$(w)$ 
can sometimes be very small over some velocity intervals and quite large 
in others.  In the interest of definiteness and uniformity, we adopted 
the following definitions for upper limits:  {\trad}$(w)$ = 0.20 
everywhere for lines that are strongly affected by ISM absorption (e.g., 
{\ion{N}{3}},{\ion{C}{3}}, and {\ion{S}{6}}); {\trad}$(w)$ = 0.15 
everywhere for lines that are modestly affected by ISM absorption or 
determined from lower resolution FOS data (e.g., {\ion{O}{6}}, 
{\ion{Si}{4}}, {\ion{C}{4}}, and {\ion{N}{5}}); and {\trad}$(w)$ = 0.10 
everywhere for weak stellar lines in {\fuse}\/ spectra that are not 
blended with ISM features (e.g., {\ion{S}{4}} and {\ion{P}{5}}).

\subsection{Ion Fractions}

The ion fractions, $q_i(w)$, were determined via Equation~(\ref{qi}) 
using the measured values of {\trad}$(w)$, {\vinf} and $\beta$ (which 
enters through the $w$ dependence), $R_\star$ (which follows from the 
observed magnitude, {\teff} and the assumed distance), the adopted 
values of $A_E$ (Table~{\ref{atomtab}}) and {\mdot} 
(Table~{\ref{startab}}). Since these introduce multiplicative errors, the 
relative uncertainties of one velocity point in $q_i(w)$ relative to 
another are the same as for {\trad}$(w)$, but the overall scaling of a 
specific $q_i(w)$ curve is affected by these errors.  Because errors in 
{\mdot}, $\beta$, $R_\star$ and {\vinf} affect all of the $q_i(w)$ curves 
for all of the ions in an individual star, they only affect comparisons of 
$q_i(w)$ among different stars.  Deviations of $A_E$ for specific 
elements from the assumed abundances will affect the relative scaling of 
the $q_i(w)$ for ions from different elements.  To determine the accuracy 
of the $q_i(w)$, we need to estimate the error in $R_\star$ in addition 
to the previously established errors.  Since $\log R_\star/R_\sun = 
-0.4 (M_V +BC -M_{bol,\sun}) -2 \log T_{{\rm eff}}/T_{{\rm eff},\sun}$, 
the major errors affecting $R_\star$ are errors in {\teff} which enter 
directly and through the $BC$ and errors in the distance modulus which 
affect $M_V$.  Using our previous estimates for these, we see that 
$R_\star$ is determined to better than 5\%.  Thus, using our previous 
results for the uncertainties in $\dot{M}$ and $v_\infty$, we see that the 
ion fractions contain multiplicative errors on the order of 25\%, which 
affect the level of an entire $q_i(w)$ curve.  Ion ratios are extremely 
useful since they are free of these multiplicative errors.  In addition, 
ratios of ion fractions from the same element are also free of assumptions 
concerning the $A_E$.

Table~\ref{fractions} lists the measured values of $q_i(w)$ for each star. 
The first column gives the normalized velocity, $w$, and the subsequent 
columns list measured values of $q_i(w)$ for each ion, with lower limits 
listed whenever the line was saturated at that velocity.  One should keep 
in mind that the {\trad}$(w)$ (and hence, the $q_i(w)$) are poorly defined 
below $w = 0.1$ due to the break-down of the Sobolev approximation, 
uncertainties in the underlying photospheric spectrum, and the assumption 
of a constant value of $w_D$.  

Table~{\ref{meanqtab}} lists the mean ion fractions, {\meanqi}, on a 
star-by-star basis.  These were calculated by integrating the ion fractions 
listed in Table~\ref{fractions} over the range $0.2 \leq w \leq 0.9$:
\begin{equation}
\langle q_i \rangle \equiv \frac{\int_{0.2}^{0.9} q_i dw}{\int_{0.2}^{0.9} 
dw} .
\end{equation}
The limits were chosen to avoid the poorly determined low velocity region 
of the fits ( $0.2 \leq w$) and the highest velocity portion of the 
profile ($w \geq 0.9$), which is often dominated by time-dependent 
phenomena like discrete absorption components 
{\citep[DACs;][]{Prinja86,Prinja87,Kaper96}}. 

We used the criteria discussed above to determine whether a mean value was 
saturated; if so, it is represented as a lower limit in 
Table~{\ref{meanqtab}}.  The few cases where {\meanqi} was determined from 
lines that are saturated only over a limited range of velocities are 
believed to be reliable estimates, and are not flagged as lower limits.
Upper limits are also indicated in Table~{\ref{meanqtab}}.  The entry 
``ISM'' for the {\ion{C}{3}} and {\ion{N}{3}} lines of {\sk{-68}{135}} 
(which is the most heavily reddened star in our sample) indicate that 
blends with the strong ISM absorptions precluded fits of the underlying 
stellar lines, which appear to be present.  These lines were excluded 
from further analysis.  The errors in the $\langle q_i \rangle$ are 
calculated using the previous results for the multiplicative error of 
25\% and noting that the point-to-point errors of 50\% in $\tau_{rad}(w)$ 
are reduced by the square root of the number of independent points that 
enter the integral, which is 15.  Together, these assumptions result in 
errors $\sim 28$\% in the {\meanqi}.  

\section{Discussion}

\subsection{Constraints on the Adopted Mass-Loss Rates}

As described above, the accuracy of any one value of $q_i(w)$ 
(Table~\ref{fractions}) is $\sim 50$\% plus 25\% uncertainties in the 
overall level of the curve, while the errors in the {\meanqi} 
(Table~\ref{meanqtab}) are $\sim 28$\%.  Since $q_i$ is inversely 
proportional to {\mdot}, and since all of the ion fractions are much less 
than unity, we can immediately conclude that the {\citeauthor{Vink00} 
\citeyearpar{Vink00,Vink01}} predictions do not systematically 
underestimate {\mdot}.  Similarly, we conclude that the adopted elemental 
abundances are probably not too small.

\subsection{Ionization Equilibria}

Table~{\ref{meanqtab}} includes measurements of {\meanqi} for multiple 
ions of C, N, and S.  In addition, we measured {\ion{P}{5}} and can 
qualitatively assess the strength of {\ion{P}{4}} as well.  With a few 
simple, though approximate, assumptions, this information can be used to 
infer details of the ionization equilibria of these elements in the stellar 
winds of O stars.

Consider first the case of carbon.  To good approximation, we expect the 
relation $q$({\ion{C}{3}}) + $q$({\ion{C}{4}}) + $q$({\ion{C}{5}}) $\approx 
1$ to hold, since the UV wind lines of {\ion{C}{2}} are not observed
in O stars, and since the ionization potential of {\ion{C}{5}} is 392 eV.  
Measurements for the resonance lines of both {\ion{C}{3}} and {\ion{C}{4}} 
are available, but are saturated in all but two cases (BI~173 and 
{\sk{-67}{101}}).  In both these cases, the combined values of 
$q$({\ion{C}{3}}) + $q$({\ion{C}{4}}) are substantially less than 1\%, 
which demonstrates that {\ion{C}{5}} is the dominant species for O stars.

For nitrogen, three O8 stars (\sk{-67}{191}, BI~173 and \sk{-67}{101}), 
have detectable and unsaturated {\ion{N}{3}} and {\ion{N}{5}} profiles.
On simplistic grounds (see, e.g., Fig.~\ref{ions}) we expect 
$q$({\ion{N}{3}}) + $q$({\ion{N}{4}}) + $q$({\ion{N}{5}}) +
$q$({\ion{N}{6}}) $\approx$ 1.  However, since {\ion{O}{6}} has a higher 
ionization potential than {\ion{N}{5}}, and in all cases $q$({\ion{O}{6}}) 
$\lesssim$0.01, we also expect $q$(\ion{N}{6}) to be $\sim$0.01.  Since 
both $q$({\ion{N}{3}}) and $q$({\ion{N}{5}}) $\lesssim 0.01$ for all three 
stars, we find that $q$({\ion{N}{4}}) $ \gtrsim$97\%.

Similarly, for S we expect $q$({\ion{S}{4}}) + $q$({\ion{S}{5}}) + 
$q$({\ion{S}{6}}) $\approx$1.  Eight stars have measurable, but unsaturated 
{\ion{S}{4}} and {\ion{S}{6}} ({\sk{-67}{166}}, {\sk{-67}{167}}, 
{\sk{-65}{22}}, {\sk{-67}{111}}, {\sk{-67}{123}}, BI 170, {\sk{-65}{214}}, 
and {\sk{-67}{05}}).  All of these indicate that $q$({\ion{S}{5}}) 
$\gtrsim 95\%$.  

For each of the preceding elements, we conclude that the dominant ion 
corresponds to the species whose resonance lines cannot be observed.
P presents an interesting counter example to this perverse situation.
Both {\ion{P}{4}} and {\ion{P}{5}} are observed in {\fuse}\/ spectra,
though the {\ion{P}{5}} doublet is present only for stars with dense winds 
owing to its low intrinsic abundance.  Once again, we assume that 
$q$({\ion{P}{4}}) + $q$({\ion{P}{5}}) + $q$({\ion{P}{6}}) $\approx$1 
and note that {\ion{P}{4}}~$\lambda$950 is very weak in all of the stars 
detected in {\ion{P}{5}} except {\sk{-65}{22}} (which has a very
low speed, dense wind).  The lack of a prominent {\ion{P}{4}} resonance 
line is significant, since its oscillator strength is nearly 2.5 times 
larger than {\ion{P}{5}} $\lambda$1117, 1128 (Table~\ref{atomtab}).  
Consequently, $q$({\ion{P}{4}}) $\ll$  $q$({\ion{P}{5}}); at most, only a 
few per cent of P is in {\ion{P}{4}}.  Furthermore, since {\ion{P}{6}} is 
produced at only slightly lower energies than {\ion{S}{6}}, we expect its 
ionization fraction to be similar.  However, Table~{\ref{meanqtab}} shows 
that $q$({\ion{S}{6}}) is never more than 2\% for stars with detectable 
{\ion{P}{5}}.  Thus, whenever it is detectable, we expect that 
{\ion{P}{5}} should be the dominant stage of ionization for P in the winds 
of O-type stars.  However, even though the observed values of 
$q$({\ion{P}{5}}) are the largest of any ion, they never exceed 25\% at 
any velocity (Table~\ref{fractions}) or 16\% when averaged over the profile 
(Table~\ref{meanqtab}).  So, although the mean ion fractions of the 
dominant, but unobserved, {\ion{N}{4}}, {\ion{C}{5}} and {\ion{S}{5}} ions 
are inferred to be greater than 0.9 in O-star winds, the measured mean ion 
fraction is less than 0.20 for the dominant {\ion{P}{5}} ion.

Of course, we realize that the arguments given above are largely heuristic, 
and that only detailed modeling can resolve the issue definitively.
Nevertheless, the unexpectedly small values derived for $q$({\ion{P}{5}})
suggest that one or more of the following systematic effects may
be responsible:

\begin{enumerate}

  \item The phosphorus abundance in the LMC is only 20-25\% of the 
        canonical value usually assumed for metals.

  \item The mass loss rates determined by the {\citeauthor{Vink00} 
        \citeyearpar{Vink00,Vink01}} relationships are 4--5 times 
        too large.   

  \item The assumed homogeneity of the stellar wind, implicit in 
        SEI modeling, is not appropriate, and causes the derived values 
        of {\trad}, and hence {\mdotqi}, to be underestimated by a 
        factor of 4--5.
       
\end{enumerate}

The first possibility is difficult to verify since there appears to be a 
lack of direct measurements of the phosphorus abundance in the LMC.  
Further, P and Al are the only abundant elements whose production is 
strongly controlled by Ne burning \citep{Anders89}.  Consequently, the 
fact that the P abundance {\em may} differ from the general abundance 
trends in the LMC might result from some detail of the Ne burning history 
of the LMC. 

Regarding the second possibility, we note that recent results  
{\citep[see, e.g.,~][]{Crowther02}}, based on the latest generation of 
wind models \citep{Hillier98} suggest that the temperature scale used by 
{\citeauthor{Vink00} \citeyearpar{Vink00,Vink01}}, {\citet{Puls96}} and 
adopted here should be revised downward.  Exactly how the revision in the 
temperature scale will affect our $q_i$ depends on how it affects the Vink 
formulae and the Puls H$\alpha$ mass loss rates.  If, e.g., the revised 
temperatures drop the rates predicted by the Vink formulae and the 
H$\alpha$ by similar amounts, then the net result is a simple relabeling 
of the temperature scale and our $q_i$ will not be affected.  If, however, 
the revision affects the Vink mass loss rates very differently than it does 
the H$\alpha$ rates, then it will have a definite impact on our results.  
An initial investigation by {\citet{Puls02}} suggests that the impact is 
minimal.  However, a definitive verdict on how the new temperature scale 
will modify our conclusions awaits further results from the new models.  

The third possibility requires some explanation.  Suppose, e.g., that 
the material of a smooth wind that is sufficiently dense to produce a 
saturated P~Cygni profile in a given ion is redistributed into 
optically thick clumps separated by transparent voids.  In contrast to the 
case of the smooth wind, the clumps only cover a fraction of the solid 
angle surrounding the star.  Consequently, the forward-scattered emission 
from this porous medium will be weaker.  Similarly, the observed absorption 
trough will not be saturated, since the face of the star is not completely 
covered by optically thick material at any velocity and unatenuated flux 
reaches the observer.  As a result, fitting a wind profile from a clumped 
wind with a homogeneous wind model will cause {\trad} to be underestimated,
and systematically low values of {\mdotq} to be derived.  Of course, 
clumping will affect the interpretation of all wind profiles, not just 
those of dominant ions, but its effects can only be unambiguously 
determined for a dominant ion.

\subsection{Ion Ratios}

As stressed earlier, ratios of the observed $q_i(w)$ are independent of 
the assumed {\mdot}, and systematic errors in $\beta$ or $R_\star$ in 
equation (\ref{qi}).  However, they can be affected by abundance 
anomalies, but these will only shift the entire curve for a specific ion 
up or down relative to an ion from another element.  These ion fraction 
ratios are interesting in two respects.  First, they show how the 
ionization of the wind changes as a function of $w$.  Second, they show 
how specific ion fractions depend on the stellar and wind parameters.  

In these contexts, two sets of ion ratios are especially important.  The 
first is ratios of different ionization stages of the same element, such 
as {\ion{S}{4}}/{\ion{S}{6}} and {\ion{N}{3}}/{\ion{N}{5}}, since these 
are also independent of assumptions about the abundances.  The second is 
ratios containing ions of S, P and Si since these non-CNO elements should 
be free of contamination by CNO processing, which can occur through the 
lifetime of an O star.  

\begin{figure}
\figurenum{6}
\epsscale{0.85}
\plotone{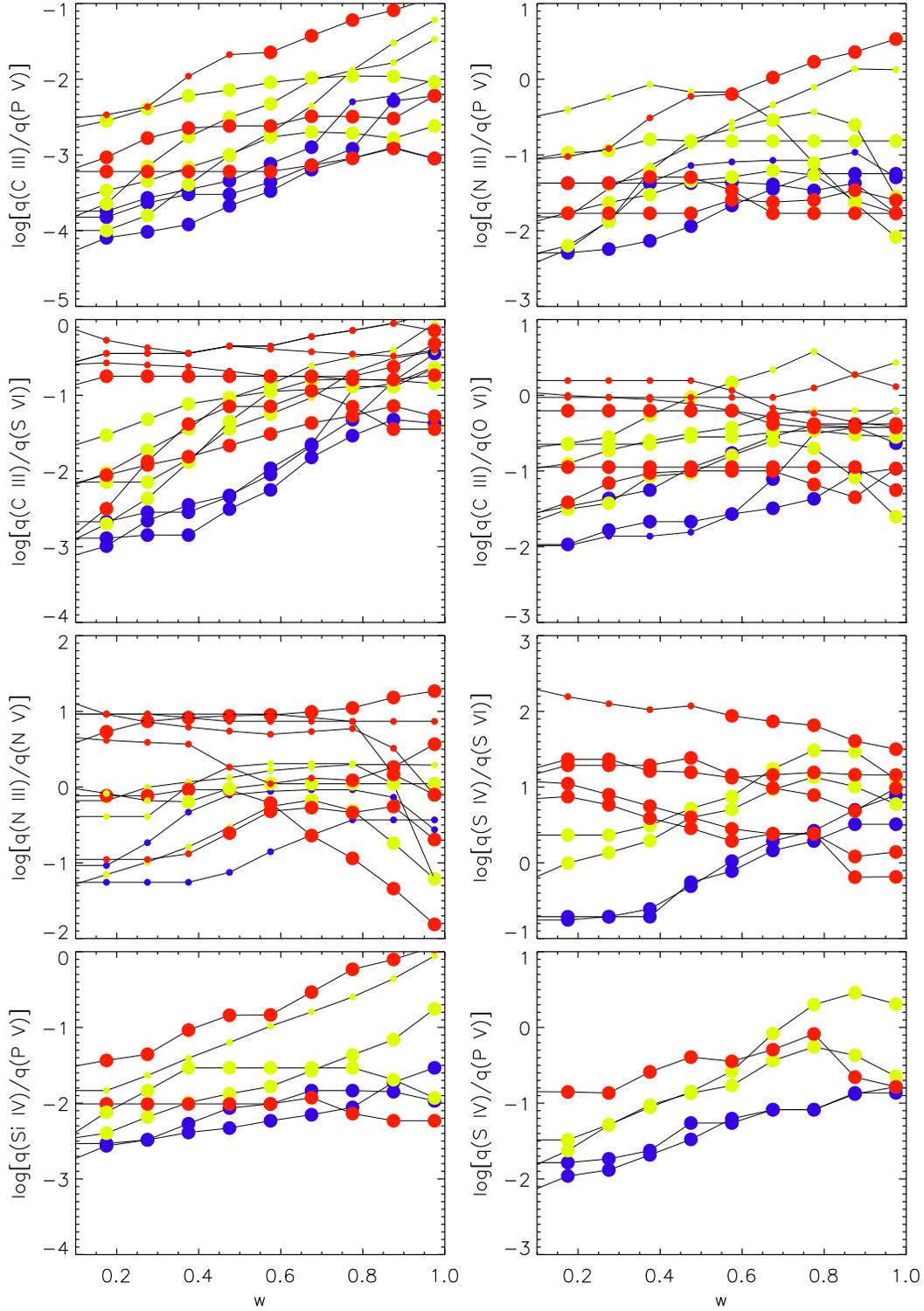}
\caption{
   Ratios of ion fractions as a function of normalized velocity, $w$, for 
   the species specified on the ordinates.  The ratios are always specified 
   with the higher ion as the denominator.  Ratios containing saturated 
   points are shown as small symbols.   Data for all stars are overplotted, 
   with different symbols representing the temperature ranges as follows --  
    red:    {\teff} $\leq$ 38~kK; 
    yellow:  38 $<$ {\teff} $\leq$ 45~kK; 
    blue:   {\teff} $>$ 45~kK.  
   \label{qratio}
}
\end{figure}

Figure~\ref{qratio} shows selected ion ratios as a function of $w$. To 
make trends easier to distinguish, adjoining velocity points were combined.  
Ratios determined from saturated $\tau_{rad}$ bins in {\em either} species 
are shown as small symbols.  The different colors represent different 
temperatures, with red for {\teff}~$\le 38$~kK, yellow for 
$38 <$~{\teff}~$\le 45$~kK, and blue for {\teff}~$> 45$ kK.  The ion with 
the higher ionization potential is always in the denominator. Since the 
winds of hotter stars should be more highly ionized, we expect their curves 
to occupy the lower portion of each figure.  

The top two plots in Figure~\ref{qratio} demonstrate that when the 
ionization potential of one ion is significantly lower than another, the 
abundance of the ion with the lower potential increases relative to the one 
with the higher potential as $w$ increases.  This result is present in all 
such ratios and is contrary to the expectations of the optically thin 
nebular approximation, which predicts the opposite effect {\citep[see, 
e.g.,~][]{Cassinelli79}}.  However, the result does agree with the 
predictions of more sophisticated models, such as those of 
\citet{Pauldrach94}.  Apparently, the ionization structure of O star winds 
is far more complex than simple approximations would lead one to believe, 
and it is hoped that analyses such as ours will provide useful constraints 
for the new generation of wind models.  

Another result can be seen in the top two panels of Figure~\ref{qratio}.  
Notice that the curves in the {\ion{C}{3}} to {\ion{P}{5}} ratio are 
reasonably well sorted with respect to {\teff} (with the cooler stars 
have relatively more of the lower ion, thus occupying the lower portion of 
the plot), while the {\ion{N}{3}} to {\ion{P}{5}} curves are not nearly as 
well sorted.  This leads to the expected result that the C abundance in the 
program stars is uniform enough for the expected temperature sort to be 
observed but the N abundance is not.  The scrambling of the ordering in the 
{\ion{N}{3}} to {\ion{P}{5}} curves is probably related to the nuclear 
history of each star.  However, precise temperatures are needed to make 
quantitative statements about the levels of N enrichment.

The second pair of plots in Figure~\ref{qratio} show another interesting 
effect.  Although the {\ion{C}{3}} to {\ion{S}{6}} ratio demonstrates the 
expected $w$ dependence for the hotter stars, the trend breaks down for the 
coolest O stars.  Similarly, the {\ion{C}{3}} to {\ion{O}{6}} ratio shows 
no $w$ dependence for any star.  The implication of this result is that the 
dominant production mechanism of {\ion{S}{6}} in the cooler O stars and 
{\ion{O}{6}} in all O stars differs from the production mechanism for the 
other elements.  Perhaps this is not surprising, since, {\ion{O}{6}} is 
{\em two stages} above the dominant stage and neither it nor the next lower 
stage can be produced by photons longward of the {\ion{He}{2}} ionization 
edge.  Consequently, it cannot even be produced by ionization of excited 
states of the dominant ion, and its production must be dominated by 
non-radiative processes, as first suggested by \citet{Cassinelli79}.  
Similarly, in the coolest O stars, the production mechanism of 
{\ion{S}{6}} has probably switched from radiative to non-radiative.

The remaining plots in Figure~\ref{qratio} provide verification for the 
previous discussion.  The {\ion{N}{3}} to {\ion{N}{5}} and {\ion{S}{4}} to 
{\ion{S}{6}} plots are for ions from the same element and free of errors in 
the assumed abundances.  For the intermediate and cooler O stars, the 
{\ion{N}{3}} to {\ion{N}{5}} ratio behaves similarly to ratios involving 
{\ion{O}{6}} or {\ion{S}{6}}, as expected.  Unfortunately, {\ion{N}{5}} is 
saturated in the hottest O stars.  The {\ion{S}{4}} to {\ion{S}{6}} plot 
verifies the {\ion{C}{3}} to {\ion{S}{6}} plot and also has ideal 
temperature sorting.  Finally, the bottom two plots show ratios of low ions 
to higher ions comprised of non-CNO species.  Once again, the $w$ 
dependence is clearly present and, with only one exception, the temperature 
ordering is as well.  The fact that the {\ion{Si}{4}} and {\ion{S}{4}} to 
{\ion{P}{5}} ratios result in excellent temperature sorting implies that 
the P abundance in the LMC is uniform, even if it is peculiar.

\subsection{Dependence of Ionization on Stellar Parameters}

\begin{figure}
\figurenum{7}
\epsscale{0.80}
\plotone{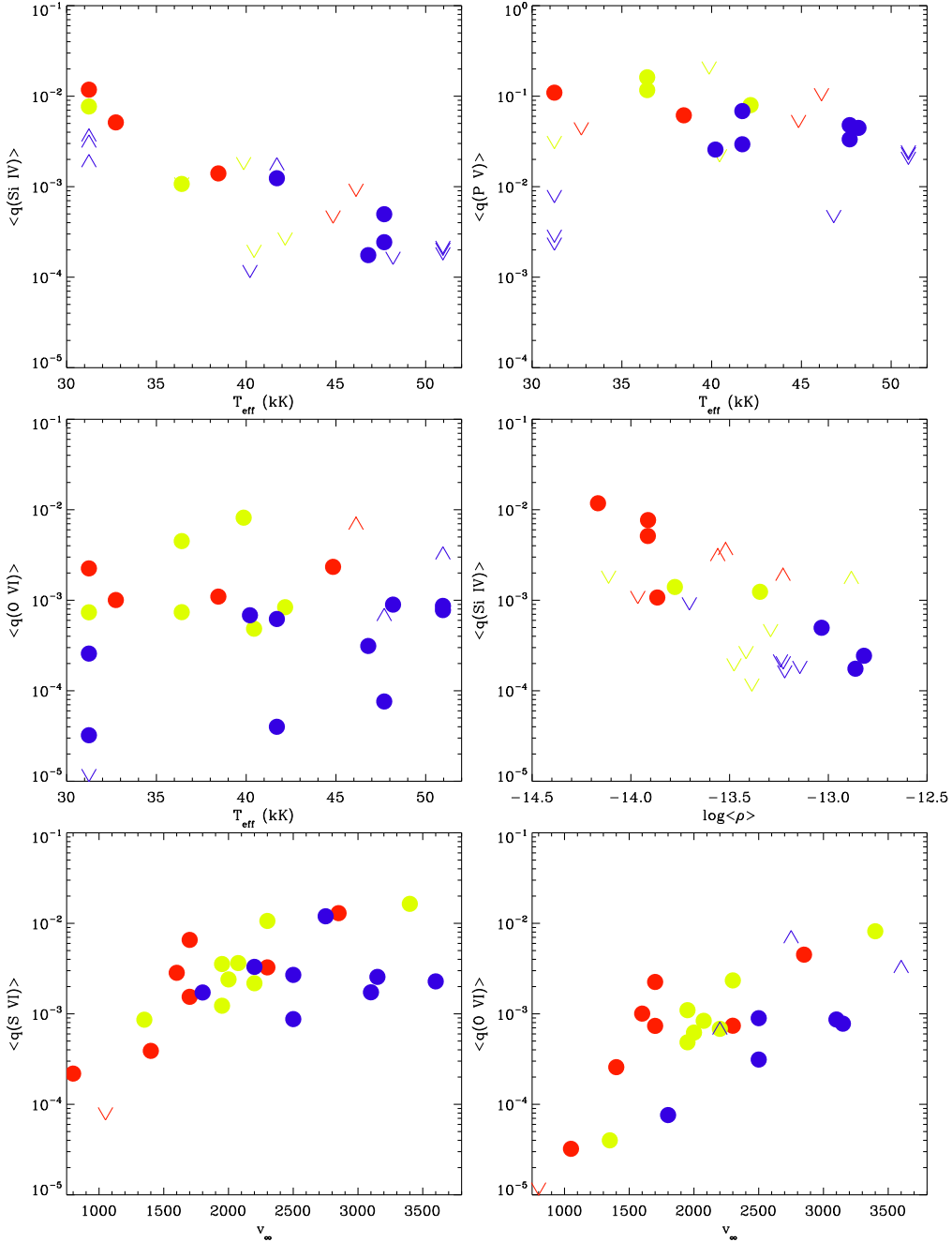}
\caption{
   Mean ion fractions of different ions plotted as functions of the stellar 
   parameters {\teff}, $\log \langle \rho \rangle$, and {\vinf}.  Different 
   symbols represent either different values of $\log \langle \rho \rangle$ 
   (for $\langle \rho \rangle$ in cgs units) or different temperatures.  
   For plots with {\teff} as the abscissa --
    red:     $\log \langle \rho \rangle \leq -13.7$;
    yellow:   $-13.7 < \log \langle \rho \rangle \leq -13.2$; 
    blue:    $-13.2 < \log \langle \rho \rangle$. 
    For plots with $\log \langle \rho \rangle$ or {\vinf} as the abscissa -- 
    red:     {\teff} $\leq$ 38~kK;
    yellow:   38 $<$ {\teff} $\leq$ 45~kK;
    blue:    45~kK $<$ {\teff}.  
   Saturated points are indicated by upward-pointing arrowheads,
   while lower limits are shown as downward-pointing arrowheads.
   \label{meanq}
}
\end{figure}

Figure~\ref{meanq} shows selected mean ionization fractions, $\langle 
q_i \rangle$, listed in Table~{\ref{meanqtab}} plotted as  functions of 
the adopted stellar parameters.  In these plots, we employ the mean density 
of the wind, which we define as 

\begin{equation}
\langle \rho \rangle \equiv \frac{\dot{M}}{4\pi v_\infty} 
\frac{\int_{0.2}^{0.9} r^{-2} v^{-1}dw }{\int_{0.2}^{0.9} dw} .
\end{equation}

Whenever a plot has anything other than {\teff} as the abscissa, the 
symbols are sorted by temperature according to the following scheme:  
red symbols for {\teff}~$\leq 38$kK yellow for $38 <$~{\teff}~$\leq 
45$kK and blue for {\teff}~$> 45$kK.  When {\teff} is the abscissa, 
then the symbols are coded according to $\log \langle \rho \rangle$ (in cgs 
units) as follows: red for $\log \langle \rho \rangle \leq -13.7$, 
yellow for $-13.7 < \log \langle \rho \rangle \leq -13.2$, and blue for 
$-13.2 < \log \langle \rho \rangle$.

The first plot, $\langle q($Si~{\sc iv}$) \rangle$ versus {\teff}, shows a  
trend which is typical of ions with low ionization potentials -- they tend 
to decrease in strength with increasing {\teff}, as expected.  The plot 
$\langle q($P~{\sc v}$) \rangle$ versus {\teff} shows that this trend 
vanishes for ions with ionization potentials equal to or greater than that 
of P~{\sc v}, implying that this may be a dominant ion.  The next plot 
shows that $\langle q($O~{\sc vi}$) \rangle$ is also independent of 
{\teff}.  Notice, however, that there is an indication that the level of 
$\langle q($O~{\sc vi}$) \rangle$ at a fixed {\teff} may decrease with 
increasing wind density.  

The $\langle q($Si~{\sc iv}$) \rangle$ versus $\log \langle \rho \rangle$ 
figure shows how plots containing limits can be deceiving.  Although there 
is an apparent trend in the symbols, upper and lower limits are present at 
the same abscissa and over wide ranges.  As a result, no strict 
relationship is present.  At first, this might seem strange, since it is 
well known that the strength of {\ion{Si}{4}} is an excellent luminosity 
discriminant and that more luminous stars tend to have denser winds.  
However, a star with a large $\langle q_i \rangle$ for an ion need not 
have a strong wind line for that ion, since the observed line strength 
depends on $\dot{M}\langle q_i \rangle$.  

Finally, the bottom panels of Figure~\ref{meanq} show the only ions which 
have a significant {\vinf} dependence.  These are the super-ions, 
{\ion{S}{6}} and {\ion{O}{6}} ({\ion{N}{5}} may as well, but it is 
difficult to be certain due to the large number of saturated points).  Once 
again, {\ion{S}{6}} seems to be a ``transition'' species in the sense that 
the ion fractions of the cooler stars (red points) have a strong dependence 
on {\vinf}, while there is effectively no correlation for the hottest 
stars (blue points).  This relation between ion fraction and {\vinf} 
suggests a mechanical origin for these ions, since {\vinf} can be viewed as 
the potential for mechanical heating through shocks.   Further, the fact 
that the ion fractions of {\ion{S}{6}} in the cooler stars and {\ion{O}{6}} 
in all stars do not depend on temperature or, equivalently, the radiation 
field of the underlying star, reinforces the idea that a non-radiative 
processes dominates their production.

\subsection{Notes on Individual Stars}

When placed in context, three stars stand out from the others.  

\medskip
\noindent{\bf BI~272.}
The wind lines of BI~272 indicate a much earlier spectral class than its 
uncertain classification of O7 II-III:.  Its terminal velocity is quite 
large (3400~\kms), which suggests it is a dwarf.  The only wind lines in 
its {\fuse}\/ spectrum are {\ion{O}{6}} (quite strong), and {\ion{S}{6}} 
(clearly present, but fairly weak; see Table~\ref{fractions}). It also has 
a strong, symmetric, {\ion{C}{3}} $\lambda$1176 feature.  Its FOS spectrum 
has a strong {\ion{N}{5}}~$\lambda\lambda$1238, 1242 wind line, but its 
{\ion{C}{4}}~$\lambda\lambda$1548,1550 line shows no sign of a wind all!  
Furthermore, the {\ion{He}{2}}~$\lambda$1640 and {\ion{N}{4}}~$\lambda$1718 
lines in the FOS spectrum show no trace of a wind, again suggesting a star 
near the main sequence.  Thus, the {\em only} wind lines in this star are 
from super-ions, and all the luminosity indicators in its spectrum point 
toward a near main sequence object.  As a result, we suspect that either 
this star is extremely hot, or either its {\fuse}\/ or FOS spectrum is 
composite.  A high-quality optical spectrum would help distinguish between 
these possibilities.

\begin{figure}
\figurenum{8}
\epsscale{0.6}
\plotone{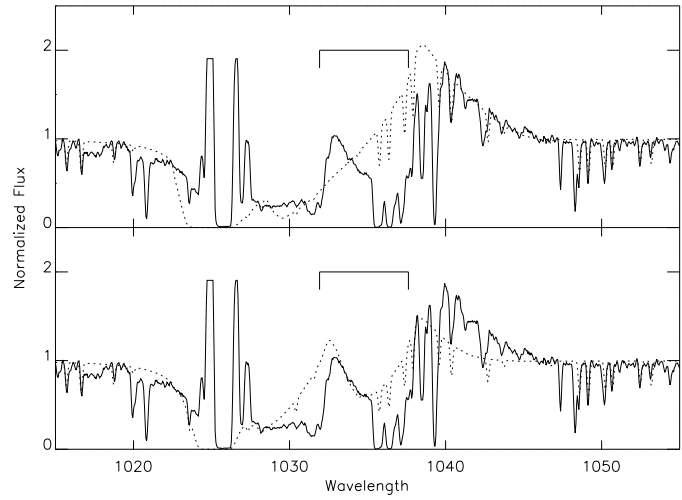}
\caption{
   Two attempts at fitting the {\ion{O}{6}} wind profile of BI~208.  
   The solid curve is normalized stellar flux, while the dashed curve is 
   the complete model with absorption from interstellar {\ion{H}{1}} and 
   H$_2$ included.  
   Both wavelength scales are in the laboratory frame.  
   The upper panel shows the best possible fit to the wind emission, 
   and the bottom panel is the best fit to the wind absorption.  
   Strong emission lines in the region of Ly~$\beta$ are due to airglow. 
   \label{bi208}
}
\end{figure}

\medskip
\noindent{\bf BI~208.}
The wind profiles of BI~208 are very peculiar.  Only the super-ions 
{\ion{O}{6}} and {\ion{S}{6}} exhibit P~Cygni profiles in {\fuse\/} 
spectra, but the {\ion{S}{6}} lines are not especially useful because 
they are weak and compromised by interstellar absorption.  Furthermore, 
the {\ion{O}{6}} profiles could not be fit self-consistently, and 
consequently are not listed in Table~\ref{fractions}.  The difficulties 
encountered with the SEI fits are illustrated in Figure~\ref{bi208}.  
The upper panel shows a fit that reproduces the emission lobe reasonably 
well with a fast ({\vinf}$\sim$2500~{\kms}), optically thick outflow, 
but which produces too much absorption.  Conversely, the lower panel 
shows that smaller values of {\trad} provide a good fit to the absorption 
trough, but fail to produce sufficient emission.  Unfortunately, the 
archival FOS spectrum is suspect, primarily because its flux distribution 
slopes downward toward shorter wavelengths to a level that is 
significantly smaller than the flux in the {\fuse}\/ spectrum at the 
same wavelength.  However, {\em if} the wind profiles in the FOS spectrum 
can be trusted, they are also quite peculiar.  For example, while 
{\ion{N}{5}} $\lambda\lambda$1238, 1240 is strong and extends to about 
$-2000$~{\kms} with substantial emission, the 
{\ion{C}{4}}~$\lambda\lambda$1548, 1550 absorption is weak ($\sim$30\% 
of the local continuum) and extends only to about $-1200$~{\kms} with no 
discernible emission.  The {\ion{N}{5}} and {\ion{C}{4}} wind profiles of 
the rapidly rotating Galactic star HD~93521 exhibit similar morphologies, 
which \citet{Massa95} and \citet{Bjorkman94} interpreted in terms of an 
outflow geometry that is cylindrically, not spherically, symmetric.  The 
similarity extends to {\ion{O}{6}}, which is also characterized by strong 
emission for comparatively weak absorption in FUV spectra of HD~93521 
obtained with the T{\"u}bingen Ultraviolet Echelle Spectrometer during 
the {\it ORFEUS-SPAS}\/ mission in 1996 {\citep{Barnstedt00}}.  However, 
in order to make this morphological connection more convincingly, a 
reliable {\hst}\/ spectrum is required.  It also would be of considerable 
interest to obtain high S/N optical spectra in order to determine whether 
BI~208 has a large {\vsini}.

\medskip
\noindent{\bf {\sk{-67}{166}}.}
The O4 supergiants {\sk{-67}{166}} and {\sk{-67}{167}} are neighbors in 
the young cluster NGC~2014.  They are separated by $\sim$11.1\arcmin\/ on 
the sky, which corresponds to a projected distance of $\sim$163~pc for 
an adopted distance modulus of 18.52.  Despite the similarity of their 
spectral types, {\citeauthor{Walborn95a} \citeyearpar{Walborn95a,
Walborn96,Walborn02a}} noted that the relative strengths of the CNO wind 
lines are completely different, with C and O weaker and N much stronger 
in {\sk{-67}{166}}.  Our quantitative analysis confirms this suspicion.  
The ratio of ion fractions in the sense {\sk{-67}{166}}\,:\,{\sk{-67}{167}} 
for the non-CNO ions {\ion{Si}{4}}, {\ion{S}{4}}, {\ion{P}{5}}, and 
{\ion{S}{6}} are 0.76, 0.73, 0.84 and 0.64, respectively, with a mean of 
0.74.  Note that these ions bracket the full range of ionization, and 
effectively give the same result.  In contrast, the three CNO lines that 
are unsaturated in at least one of the stars --  {\ion{C}{3}}, 
{\ion{N}{3}}, and {\ion{O}{6}} -- have ratios of  0.20, $\geq$1 and 
$\leq$0.10, respectively; i.e., the abundances of carbon and oxygen 
in {\sk{-67}{166}} are several times less than in {\sk{-67}{167}} and its 
nitrogen abundance is greater.  This pattern of abundances is a signature 
of material processed through the CNO-cycle and implies that more 
processed material is present on the surface of {\sk{-67}{166}} than 
{\sk{-67}{167}}; i.e., that {\sk{-67}{166}} is an ON star.  Unfortunately, 
the spectral properties of these stars do match closely enough to permit 
the more quantitative differential abundance analysis used by 
\citet{Massa91}.  However, detailed modeling by \citet{Crowther02} 
confirms that N is enhanced and C and O are depleted in the atmosphere 
{\sk{-67}{166}} compared to normal LMC abundances.

\section{Summary and Conclusions}

Using the unique spectral coverage and sensitivity provided by {\fuse}\/ 
and SEI modelling to translate the observed wind line profiles into 
quantitative information, we determined the following:

\begin{enumerate}

\item Because none of the derived ion fractions exceed unity, the mass 
      loss rates determined by the {\citeauthor{Vink00} 
      \citeyearpar{Vink00,Vink01}} formulae are {\em not} too small.

\item Because the {\ion{P}{5}} ion fraction never approaches unity, as 
      expected by models, we conclude that either 
  \begin{enumerate}
      \item the phosphorus abundance in the LMC is $\sim 1/3$ of the 
            canonical value assumed for most elements, or 
      \item the Vink et al.\ mass loss rates are 2 to 3 times too large, or 
      \item some aspect of the models, such as their neglect of clumping in 
            the winds, results in derived {\mdotqi} values that are $\sim 
            3$ times too small. 
  \end{enumerate}

\item Ion ratios not involving N ions show much clearer {\teff} dependence 
      (with lower ions being more dominant in the cooler stars) than do 
      ion ratios involving N ions.  This result is attributed to a 
      non-uniformity in the abundance of nitrogen, as a result of nuclear 
      processing over the lifetimes of the stars.

\item Ion fractions of higher ions, that are not super-ions. decrease 
      relative to lower ions as $w=v/${\vinf} increases.

\item Ionic ratios containing super-ions do not show a $w$ dependence.  
      {\ion{O}{6}} exhibits this trait for all O stars, {\ion{S}{6}} shows 
      it for only the coolest O stars, and the {\ion{N}{5}} lines are too 
      often saturated to distinguish between the early and late O stars.  
      Together, this result implies that the production of {\ion{O}{6}} 
      is non-radiative, and the same is true for {\ion{S}{6}} in the 
      coolest O stars.

\item The mean ion fractions, $\langle q(${\ion{S}{6}}$)\rangle$ and 
      $\langle q(${\ion{O}{6}}$)\rangle$, do not depend on the temperature 
      of the star, again suggesting a non-radiative production mechanism,

\item The ion fractions of {\ion{O}{6}} in all stars and {\ion{S}{6}} in 
      the later O stars are positively correlated with {\vinf}, suggesting 
      a shock strength dependence.

\item The wind lines in BI~272 are indicative of a considerably hotter star 
      than is implied by its uncertain spectral.

\item BI~208 may have a non-spherical wind.

\item {\sk{-67}{166}} is an ON star. 

\end{enumerate}

\acknowledgements
We are indebted to the {\fuse}\/ science team for making the data 
available to us, to Henny Lamers for stimulating discussion, to Jorick Vink 
for a critical reading for the manuscript and to Joe Casinelli for a rapid 
and thorough referee report.  {\fuse} is operated for NASA by the Johns 
Hopkins University under NASA contract NAS5-32985.  Some of the data 
presented in this paper were obtained from the Multimission Archive at the 
Space Telescope Science Institute (MAST). STScI is operated by AURA, under 
NASA contract NAS5-26555 and support for non-{\it HST}\/ data is provided 
in part by NASA OSS grant NAG5-7584.

\clearpage
\clearpage

\clearpage

\clearpage

\clearpage

\clearpage

\clearpage
\begin{deluxetable}{lllcccrrc}
\tabletypesize{\footnotesize}
\tablecolumns{8}
\tablewidth{0pt}
\tablecaption{Fundamental Properties of Targets\label{startab}}
\tablehead{ \colhead{Name}                                               &
            \colhead{Alias}                                              &
            \colhead{Spectral Type}                                      &
            \colhead{Ref.\,\tablenotemark{a}}                            &
            \colhead{\vinf}                                              &
            \colhead{$\log L_\star/\,{\rm L}_{\sun}$\,\tablenotemark{b}} & 
            \colhead{\teff \,\tablenotemark{c}}                          &  
            \colhead{$R_\star$}                                          &
            \colhead{\mdot\,\tablenotemark{d}}                          \\
            \colhead{}                                                   &
            \colhead{}                                                   &  
            \colhead{}                                                   &
            \colhead{}                                                   &
            \colhead{(\kms)}                                             &
            \colhead{}                                                   &
            \colhead{(K)}                                               &
            \colhead{(R$_\sun$)}                                         &
            \colhead{($10^{-6}$ \msunpyr)}                               }
\startdata

\sk{-67}{211} & HDE 269810 & O2 III(f$^\ast$)    & 1 & 3600   & 6.32 & 50960 & 18.5 & 10.5 \phn \\
\sk{-66}{172} &            & O2 III(f$^\ast$)+OB & 1 & 3100   & 6.12 & 50960 & 14.7 & \phn 5.92 \\
\sk{-70}{ 91} &            & O2 III(f$^\ast$)+OB & 2 & 3150   & 6.12 & 50960 & 14.7 & \phn 5.82 \\ 
\sk{-67}{166} & HDE 269698 & O4 If+              & 3 & 1800   & 6.26 & 47690 & 19.7 & 15.9 \phn \\
\sk{-67}{167} &            & O4 Inf+             & 4 & 2200   & 6.19 & 47690 & 18.2 & 10.1 \phn \\ \\
\sk{-67}{ 69} &            & O4 III(f)           & 4 & 2500   & 6.02 & 48180 & 14.7 & \phn 4.83 \\
\sk{-71}{ 45} & HDE 269676 & O4-5 III(f)         & 3 & 2500   & 6.58 & 46800 & 29.6 & 35.8 \phn \\
\enddata

\tablenotetext{a}{References for Spectral Type.~
    1 = \citealt{Walborn02b};
    2 = \citealt{Walborn02a};
    3 = \citealt{Walborn77};
    4 = \citealt{Garmany87a};
    5 = \citealt{Walborn95a};
    6 = Preliminary classification by N. R. Walborn, private communication;
    7 = \citealt{Conti86};  
    8 = \citealt{Fitzpatrick88}.
    }

\tablenotetext{b}{{$\log L/L_\sun$} is based on a distance modulus to the
                   LMC of 18.52 from \citealt{Fitz2002}.}

\tablenotetext{c}{From \citealt{Vacca96}.}

\tablenotetext{d}{From \citealt{Vink01}.}

\end{deluxetable}
\begin{deluxetable}{lclcccll}
\tabletypesize{\small}
\tabletypesize{\footnotesize}
\tablecolumns{8}
\tablewidth{0pt}
\tablecaption{Sources of Photometric and Spectroscopic Data\label{datatab}}
\tablehead{ \colhead{Name}                                     &
            \colhead{$V$}                                      &
            \colhead{(\bv)}                                    &
            \colhead{Ref.\,\tablenotemark{a}}                  &
            \colhead{\ebv}                                     &
            \colhead{$F_\lambda$(1150 \AA)\,\tablenotemark{b}} &
            \colhead{\fuse}~~~~~~~~~~~~~                       &
            \colhead{{\it HST}/FOS, {\it HST}/STIS, {\it IUE}}  }

\startdata
\sk{-67}{211} & 12.28 & $-$0.23  & 1 & 0.09 &    33.7 & P1171603       & FOS: Y14M0D05T              \\
\sk{-66}{172} & 13.13 & $-$0.12  & 2 & 0.20 &    10.7 & P1172201       & FOS: Y0YG0903T, 4T, 5T      \\
\sk{-70}{91}  & 12.78 & $-$0.23  & 3 & 0.09 &    17.0 & P1172501       & FOS: Y25U0L01T              \\
\sk{-67}{166} & 12.27 & $-$0.22  & 1 & 0.10 &    28.4 & A1330101 -- 62 & {\iue}: SWP\,6967           \\
\sk{-67}{167} & 12.54 & $-$0.19  & 4 & 0.13 &    21.9 & P1171901, 2    & FOS: Y14M0C09T              \\  \\
\sk{-67}{69}  & 13.09 & $-$0.16  & 3 & 0.16 &    10.4 & P1171703       & \multicolumn{1}{c}{\nodata} \\
\sk{-71}{45}  & 11.51 & $-$0.19  & 5 & 0.13 &    26.3 & P1031501 -- 4  & STIS: O63521010             \\
\enddata

\tablenotetext{a}{References for optical photometry.~
                  1 = \citealt{Ardeberg72};
                  2 = \citealt{Rousseau78};
                  3 = \citealt{Isserstedt79};
                  4 = \citealt{Isserstedt75};
                  5 = \citealt{Heydari91};
                  6 = \citealt{Isserstedt82};
                  7 = \citealt{Brunet75};
                  8 = \citealt{Fitzpatrick88}.
                  }
                  
\tablenotetext{b}{Flux observed by {\fuse} at 1150 {\AA} in units of $10^{-13}$ erg/s/cm$^2$/{\AA}.}

\end{deluxetable}

\begin{deluxetable}{lccrrcll}
\tabletypesize{\small}
\tablecolumns{8}
\tablewidth{0pt}
\tablecaption{Properties of FUV and UV Resonance Lines\label{atomtab}}
\tablehead{ \colhead{Ion}                                     &
            \colhead{Abundance\,\tablenotemark{a}}            &
            \colhead{I.P. Range\,\tablenotemark{b}}           &
            \colhead{$\lambda_{\rm blue}$\,\tablenotemark{c}} &
            \colhead{$\lambda_{\rm red} $\,\tablenotemark{d}} &
            \colhead{$\Delta v$\,\tablenotemark{e}}           &
            \colhead{$f_{\rm blue}$\,\tablenotemark{f}}       &
            \colhead{$f_{\rm red} $\,\tablenotemark{g}}       \\
            \colhead{}                                        &
            \colhead{}                                        &
            \colhead{(eV)}                                    &
            \colhead{(\AA)}                                   &
            \colhead{(\AA)}                                   &
            \colhead{(\kms)}                                  &
            \colhead{}                                        &
            \colhead{}                                         }

\startdata

\ion{C}{3}  & 8.55 &    24.383\,--\,47.887  &  977.020 & \multicolumn{1}{c}{\nodata} & \nodata & 0.759  & \multicolumn{1}{c}{\nodata} \\
\ion{C}{4}  & 8.55 &    47.887\,--\,64.492  & 1548.203 & 1550.777                    & \phn498 & 0.191  & 0.0952                      \\ \\
\ion{N}{3}  & 7.97 &    29.601\,--\,47.448  &  989.799 &  991.571\,\tablenotemark{h} & \phn537 & 0.123  & 0.122\,\tablenotemark{i}    \\
\ion{N}{5}  & 7.97 &    77.472\,--\,97.888  & 1238.821 & 1252.804                    & \phn964 & 0.156  & 0.078                       \\ \\
\ion{O}{6}  & 8.87 &   113.896\,--\,138.116 & 1031.926 & 1037.617                    &    1653 & 0.133  & 0.0659                      \\ \\
\enddata

\tablenotetext{a}{Solar-system abundances by number from the solar 
                  photosphere {\citep[C, N, O;][]{Grevesse93}} or 
                  meteorites {\citep[Si, P, S;][]{Anders89}}, expressed as  
                  $\log N / N_H  + 12.00$.
                  These values were uniformly scaled by 0.5 in order to 
                  obtain the abundances adopted for the LMC.}
                  
\tablenotetext{b}{The lower and upper ionization potentials for the ion 
                  from \citealt{Moore70}.}

\tablenotetext{c}{Rest wavelength of the blue component of the doublet 
                  from \citealt{Morton91} or \citealt{Morton02}.}

\tablenotetext{d}{Rest wavelength of the red component of the doublet 
                  from \citealt{Morton91} or \citealt{Morton02}.}                 

\tablenotetext{e}{Separation of the doublet components expressed in {\kms}.}

\tablenotetext{f}{Oscillator strength for the blue component of the 
                  doublet from \citealt{Morton91} or \citealt{Morton02}.}

\tablenotetext{g}{Oscillator strength for the red component of the doublet
                  from \citealt{Morton91} or \citealt{Morton02}.}
                  
\tablenotetext{h}{The mean wavelength from two low-lying excited transitions 
                  connected to the ground state, weighted by their respective
                  oscillator strengths.}

\tablenotetext{i}{The sum for two low-lying transitions connected to the 
                  ground state, one of which is $\sim$10 times stronger than 
                  the other.}             

\end{deluxetable}
\begin{deluxetable}{lcccccccccccccccccc}
\tabletypesize{\small}
\tablecolumns{9}
\tablewidth{0pt}
\tablecaption{Parameters of Photospheric Features\label{phototab}}
\tablehead{ \colhead{Name              } &
            \colhead{\ion{S}{6}        } &
            \colhead{\ion{N}{5}        } &
            \colhead{\ion{C}{4}        } &
            \colhead{\ion{P}{5}        } &
            \colhead{\ion{S}{4}        } &
            \colhead{\ion{Si}{4}       } &
            \colhead{\ion{C}{3}        } &
            \colhead{\ion{N}{3}        }   }

\startdata

\sk{-67}{211} & 3.0    ,    200   &  
                0.0    ,\phn\phn0 & 
                8.1    ,    300   &
               \nodata            &
               \nodata            &  
               \nodata            &
               \nodata            &
               \nodata              \\
\sk{-66}{172} & 0.0    ,\phn\phn0 & 
                2.0    ,    250   & 
                2.0    ,    250   &
               \nodata            &
               \nodata            &  
               \nodata            &
               \nodata            &
               \nodata              \\
\sk{-70}{91}  & 0.0    ,\phn\phn0 &
                0.0    ,\phn\phn0 & 
                4.0    ,    250   &
               \nodata            &
               \nodata            &
               \nodata            &
               \nodata            &
               \nodata             \\
\sk{-67}{166} & 3.0    ,    200   & 
                1.5    ,    149   & 
                1.0    ,    101   & 
                2.0    ,    200   & 
                0.0    ,\phn\phn0 & 
                3.0    ,    300   & 
                2.0    ,    200   & 
                6.0    ,    200     \\
\sk{-67}{167} & 5.0    ,    200   & 
                0.0    ,\phn\phn0 & 
                1.0    ,    101   & 
                2.0    ,    300   & 
                0.5    ,     98   & 
                1.0    ,    400   & 
                0.2    ,    101   &  
                2.0    ,    150     \\  \\
\sk{-67}{69}  & 2.0    ,    300   &
               \nodata            &
               \nodata            & 
                2.0    ,    300   &
               \nodata            &
               \nodata            &  
                2.0    ,    300   & 
                2.0    ,    300     \\
\enddata

\tablecomments{
      The entries give the adopted values of $(\tau_0^B, v_G)$, with
      $v_G$ in {\kms}.  
      Zeros indicate a flat continuum;
      ellipsis indicate that the line was not fit.
      Flat continua were adopted for the {\ion{O}{6}} lines, with
      two exceptions: 
      (a) for {\sk{-67}{167}} we used $(\tau_0^B, v_G) = (4.9,~200)$;
      and 
      (b) for {\sk{-66}{169}} the {\ion{O}{6}} wind lines were not fit.
      }

\end{deluxetable}

\begin{deluxetable}{lcrrrrrrrrrr}
\tabletypesize{\small}
\tablecolumns{12}
\tablewidth{0pt}
\tablecaption{Measured Parameters of the Wind Velocity Law\label{veltab}}

\tablehead{ \colhead{Name                            }  &
            \colhead{\vinf                           }  &
            \colhead{$\beta$                         }  &
            \colhead{                                }  &           
            \colhead{                                }  &           
            \colhead{                                }  &
            \colhead{                                }  &
            \colhead{$w_D$\,\tablenotemark{a}        }  &       
            \colhead{                                }  &
            \colhead{                                }  &           
            \colhead{                                }  &
            \colhead{                                }  \\
            \colhead{                                }  &       
            \colhead{(\kms)                          }  &           
            \colhead{                                }  &           
            \colhead{ \ion{O}{6}                     }  &
            \colhead{ \ion{S}{6}                     }  &
            \colhead{ \ion{N}{5}                     }  &
            \colhead{ \ion{C}{4}                     }  &
            \colhead{ \ion{P}{5}                     }  &
            \colhead{ \ion{S}{4}                     }  &
            \colhead{ \ion{Si}{4}                    }  &
            \colhead{ \ion{C}{3}                     }  &
            \colhead{ \ion{N}{3}                     }    }

\startdata
\sk{-67}{211} & 3600   & 0.8 & 0.06  & 0.06  & 0.06  & 0.06  &\nodata&\nodata&\nodata&\nodata&\nodata\\
\sk{-66}{172} & 3100   & 1.0 & 0.08  & 0.08  & 0.10  & 0.08  &\nodata&\nodata&\nodata&\nodata&\nodata\\
\sk{-70}{91}  & 3150   & 1.0 & 0.05  & 0.06  & 0.10  & 0.06  &\nodata&\nodata&\nodata&\nodata&\nodata\\
\sk{-67}{166} & 1800   & 1.0 & 0.15  & 0.15  & 0.15  & 0.15  & 0.15  & 0.15  & 0.15  & 0.15  & 0.15  \\
\sk{-67}{167} & 2200   & 1.0 & 0.15  & 0.15  & 0.20  & 0.15  & 0.10  & 0.10  & 0.10  & 0.10  & 0.10  \\ \\
\sk{-67}{69}  & 2500   & 1.0 & 0.08  & 0.08  &\nodata&\nodata& 0.08  &\nodata&\nodata& 0.08  & 0.08  \\
\sk{-71}{45}  & 2500   & 0.8 & 0.08  & 0.08  & 0.08  & 0.08  &\nodata&\nodata& 0.08  &\nodata&\nodata\\
\enddata

\tablenotetext{a}{Turbulent velocity, expressed in units of {\vinf}.}

\end{deluxetable}

\begin{deluxetable}{lccccc}
\tablecolumns{6}
\tablewidth{0pt}
\tablecaption{Comparison with Previous Determinations of 
               the Velocity Law\label{cfvlaw}} 
\tablehead{ \colhead{Name}                     &
            \colhead{Present Work}             &
	    \colhead{Haser\,\tablenotemark{a}} &
	    \colhead{PP\,\tablenotemark{b}   } &
            \colhead{PC\,\tablenotemark{c}   } &     
	    \colhead{Other}                     \\
	    \colhead{}                         &
	    \colhead{\vinf, $\beta$}           &	
	    \colhead{\vinf, $\beta$}           &
	    \colhead{\vinf, $\beta$}           &
	    \colhead{\vinf}                    &
	    \colhead{\vinf}                    }
            
\startdata

\sk{-67}{211} & 3600, 0.8 & 3750, 0.55 & \nodata   & \nodata & 3800\,\tablenotemark{d} \\
\sk{-66}{172} & 3100, 1.0 & 3250, 0.70 & \nodata   & 2950    & \nodata                 \\
\sk{-67}{166} & 1800, 1.0 & 1900, 0.70 & 2250, 1.1 & 1735    & 2020\,\tablenotemark{d} \\
\sk{-67}{167} & 2200, 1.0 & 2150, 0.70 & \nodata   & 2005    & \nodata                 \\
\sk{-70}{ 69} & 2750, 1.0 & 2600, 1.00 & \nodata   & \nodata & \nodata                 \\ \\
\sk{-70}{ 60} & 2300, 1.0 & \nodata    & \nodata   & 2150    & \nodata                 \\
\sk{-67}{111} & 2000, 1.0 & \nodata    & 2090, 0.7 & \nodata & 1800\,\tablenotemark{e} \\
\enddata
\tablecomments{All values of {\vinf} are in {\kms}.} 
\tablenotetext{a}{\citealt{Haser95}; based on fits with an independent
                  wind-profile fitting technique.}
\tablenotetext{b}{\citealt{Patriarchi92}; based on fits with the SEI program
                  of \citealt{Lamers87}.}
\tablenotetext{c}{\citealt{Prinja98}; based measurements of the position of narrow
                   absorption components or the most blueshifted part of an extended
		   region of saturation (a ``black trough'').} 
\tablenotetext{d}{\citealt{Bernabeu89}; based on measurement of the maximum
                   velocity seen in absorption, usually called {$v_{\rm edge}$}.
		   This is now generally interpreted to be an overestimate of {\vinf}.}
\tablenotetext{e}{\citealt{Bianchi00}; based on fits with the 
                   SEI program of \citealt{Lamers87}. A value of
		   $\beta$ = 1 was derived.}		   		   

\end{deluxetable}
\begin{deluxetable}{crrrrrrrrr}
\tabletypesize{\small}
\tablecolumns{10}
\tablewidth{0pt}
\tablecaption{Measured Ion Fractions\label{fractions}}
\tablehead{ \colhead{$v/v_\infty$} &
\colhead{ \ion{C}{3} } &
\colhead{ \ion{N}{3} } &
\colhead{ \ion{Si}{4}} &
\colhead{ \ion{S}{4} } &
\colhead{ \ion{C}{4} } &
\colhead{ \ion{P}{5} } &
\colhead{ \ion{N}{5} } &
\colhead{ \ion{S}{6} } &
\colhead{ \ion{O}{6} } }
\startdata
\multicolumn{10}{c}{} \\
\multicolumn{10}{c}{\sk{-67}{211}} \\
\hline \\
0.01 & \multicolumn{1}{c}{\nodata} & \multicolumn{1}{c}{\nodata} &
\multicolumn{1}{c}{\nodata} & \multicolumn{1}{c}{\nodata} & $ 2.1~(-5)$ &
\multicolumn{1}{c}{\nodata} & $>7~(-3)$ & $ 1.4~(-3)$ & $>4~(-4)$ \\ 0.05 &
\multicolumn{1}{c}{\nodata} & \multicolumn{1}{c}{\nodata} &
\multicolumn{1}{c}{\nodata} & \multicolumn{1}{c}{\nodata} & $ 4.2~(-5)$ &
\multicolumn{1}{c}{\nodata} & $>1~(-2)$ & $ 2.9~(-3)$ & $>8~(-4)$ \\ 0.10 &
\multicolumn{1}{c}{\nodata} & \multicolumn{1}{c}{\nodata} &
\multicolumn{1}{c}{\nodata} & \multicolumn{1}{c}{\nodata} & $ 6.8~(-5)$ &
\multicolumn{1}{c}{\nodata} & $>2~(-2)$ & $ 4.7~(-3)$ & $>1~(-3)$ \\ 0.15 &
\multicolumn{1}{c}{\nodata} & \multicolumn{1}{c}{\nodata} &
\multicolumn{1}{c}{\nodata} & \multicolumn{1}{c}{\nodata} & $ 9.1~(-5)$ &
\multicolumn{1}{c}{\nodata} & $>3~(-2)$ & $ 5.1~(-3)$ & $>1~(-3)$ \\ 0.20 &
\multicolumn{1}{c}{\nodata} & \multicolumn{1}{c}{\nodata} &
\multicolumn{1}{c}{\nodata} & \multicolumn{1}{c}{\nodata} & $ 1.1~(-4)$ &
\multicolumn{1}{c}{\nodata} & $>4~(-2)$ & $ 4.7~(-3)$ & $>2~(-3)$ \\
\enddata
\tablecomments{
The format n.n~(E) represents the number
n.n~$\times~10^{{\rm E}}$. The ion fractions
were computed with the assumption that
$A_E$(LMC) = 0.50\,$A_E$(Galaxy). }
\end{deluxetable}
\begin{deluxetable}{lrrrrrrrrr}
\tabletypesize{\footnotesize}
\tablecolumns{10}
\tablewidth{0pt}
\tablecaption{Mean Ion Fractions\label{meanqtab}}
\tablehead{ \colhead{Name         }  &
            \colhead{ \ion{C}{3}  }  &
            \colhead{ \ion{N}{3}  }  &
            \colhead{ \ion{Si}{4} }  &
            \colhead{ \ion{S}{4}  }  &
            \colhead{ \ion{C}{4}  }  &
            \colhead{ \ion{P}{5}  }  &
            \colhead{ \ion{S}{6}  }  &
            \colhead{ \ion{N}{5}  }  &
            \colhead{ \ion{O}{6}  }  } 

\startdata
\sk{-67}{211} & $ <2  ~(-5)$                 & 
                $ <6  ~(-4)$                 & 
                $ 1.5 ~(-4)$                 & 
                $ <2  ~(-3)$                 & 
                $ 3.8 ~(-4)$                 & 
                $ <1  ~(-2)$                 &   
                $ 2.2 ~(-3)$                 & 
                $ >3  ~(-2)$                 & 
                $ >3  ~(-3)$                 \\
\sk{-66}{172} & $<2   ~(-5)$                 &
                $ <6  ~(-4)$                 &
                $ 1.7 ~(-4)$                 &
                $<3   ~(-3)$                 &
                $3.3  ~(-4)$                 &
                $ <2  ~(-2)$                 &
                $ 1.7 ~(-3)$                 &
                $>1   ~(-2)$                 &
                $ 8.6 ~(-4)$                 \\
\sk{-70}{91}  & $<2   ~(-5)$                 &
                $<7   ~(-4)$                 &   
                $ 1.8 ~(-4)$                 &     
                $<3   ~(-3)$                 &   
                $ 2.3 ~(-4)$                 &     
                $<2   ~(-2)$                 &   
                $ 2.5 ~(-3)$                 &   
                $ 2.9 ~(-3)$                 &   
                $ 7.7 ~(-4)$                 \\
\sk{-67}{166} & $ 1.7 ~(-5)$                 &     
                $>2   ~(-3)$                 &   
                $ 2.4 ~(-4)$                 &   
                $ 1.9 ~(-3)$                 &     
                $>3   ~(-4)$                 &   
                $ 3.3 ~(-2)$                 &   
                $ 1.7 ~(-3)$                 &     
                $>3   ~(-3)$                 &   
                $ 7.6 ~(-5)$                 \\
\enddata

\tablecomments{
     The format n.n~(E) represents the number n.n~$\times~10^{{\rm E}}$.  
     The ion fractions for all species were computed with the assumption 
     that $A_E$(LMC) =  0.50\,$A_E$(Galaxy).  }
      
\tablenotetext{a}{The observed line is dominated by an interstellar
                  component.}

\end{deluxetable}

\end{document}